\documentclass[12pt,oneside,reqno]{amsart}
\usepackage{amsmath,amssymb,amsthm,textcomp}
\usepackage{amsfonts,graphicx}
\usepackage[mathscr]{eucal}
\usepackage{booktabs}
\usepackage{siunitx}

\usepackage{subfig}
\usepackage{epsfig}
\usepackage{caption}
\theoremstyle{definition}
\usepackage{float}
\usepackage[colorlinks=true,urlcolor=blue,citecolor=red,linkcolor=blue,bookmarks=true]{hyperref}
\usepackage{rotating,tabularx}
\usepackage{lscape}
\usepackage[table,xcdraw]{xcolor}
\usepackage{multirow}
\usepackage[inline]{enumitem}
\floatstyle{plaintop}
\restylefloat{table}
\numberwithin{equation}{section}
\usepackage{epstopdf}
\usepackage{booktabs}

\newcommand{\ncom}{\newcommand}

\ncom{\beq}{\begin{equation}}
	\ncom{\eeq}{\end{equation}}
\ncom{\bea}{\begin{eqnarray*}}
	\ncom{\eea}{\end{eqnarray*}}
\ncom{\beqa}{\begin{eqnarray}}
	\ncom{\eeqa}{\end{eqnarray}}
\ncom{\nno}{\nonumber}
\ncom{\non}{\nonumber}
\ncom{\ds}{\displaystyle}
\ncom{\half}{\frac{1}{2}}
\ncom{\mbx}{\makebox{.25cm}}
\ncom{\hs}{\mbox{\hspace{.25cm}}}
\ncom{\rar}{\rightarrow}
\ncom{\Rar}{\Rightarrow}
\ncom{\noin}{\noindent}
\ncom{\bc}{\begin{center}}
	\ncom{\ec}{\end{center}}
\ncom{\sz}{\scriptsize}
\ncom{\rf}{\ref}
\ncom{\s}{\sqrt{2}}
\ncom{\sgm}{\sigma}
\ncom{\Sgm}{\Sigma}
\ncom{\psgm}{\sigma^{\prime}}
\ncom{\dt}{\delta}
\ncom{\Dt}{\Delta}
\ncom{\lmd}{\lambda}
\ncom{\Lmd}{\Lambda}
\ncom{\Th}{\Theta}
\ncom{\e}{\eta}
\ncom{\eps}{\epsilon}
\ncom{\pcc}{\stackrel{P}{>}}
\ncom{\lp}{\stackrel{L_{p}}{>}}
\ncom{\dist}{{\rm\,dist}}
\ncom{\sspan}{{\rm\,span}}
\ncom{\re}{{\rm Re\,}}
\ncom{\im}{{\rm Im\,}}
\ncom{\sgn}{{\rm sgn\,}}
\ncom{\ba}{\begin{array}}
	\ncom{\ea}{\end{array}}
\ncom{\hone}{\mbox{\hspace{1em}}}
\ncom{\htwo}{\mbox{\hspace{2em}}}
\ncom{\hthree}{\mbox{\hspace{3em}}}
\ncom{\hfour}{\mbox{\hspace{4em}}}
\ncom{\vone}{\vskip 2ex}
\ncom{\vtwo}{\vskip 4ex}
\ncom{\vonee}{\vskip 1.5ex}
\ncom{\vthree}{\vskip 6ex}
\ncom{\vfour}{\vspace*{8ex}}
\ncom{\norm}{\|\;\;\|}
\ncom{\integ}[4]{\int_{#1}^{#2}\,{#3}\,d{#4}}
\ncom{\vspan}[1]{{{\rm\,span}\{ #1 \}}}
\ncom{\dm}[1]{ {\displaystyle{#1} } }
\ncom{\ri}[1]{{#1} \index{#1}}

\newtheorem{remark}{\bf Remark}[section]
\newtheorem{proposition}{Proposition}[section]
\newtheorem{lemma}{Lemma}[section]
\newtheorem{corollary}{Corollary}[section]

\newtheoremstyle
{remarkstyle}
{}
{11pt}
{}
{}
{\bfseries}
{:}
{     }
{\thmname{#1} \thmnumber{#2} }

\theoremstyle{remarkstyle}

\def\eps{\varepsilon}

\makeatletter
\@namedef{subjclassname@2020}{%
	\textup{2020} Mathematics Subject Classification}
\makeatother
\begin{document}
	%\title{\large A\lowercase{ study of dependence measures and parametric inference of copula-based regression model for bivariate Weibull distribution.}}
	\title{\large C\lowercase{opula-}\large B\lowercase{ased} \large E\lowercase{stimation of} \large C\lowercase{ausal} \large E\lowercase{ffects in} \large M\lowercase{ultiple} \large L\lowercase{inear and} \large P\lowercase{ath} \large A\lowercase{nalysis} M\lowercase{odels} }
	%\title{\large E\lowercase{stimation of 
			%	 Effects in Path Analysis:} \large A\lowercase{ copula approach}}
	\author[Alam Ali]{Alam Ali$^{1}$}
	\author{Ashok Kumar Pathak$^{1*}$}
	\author{ Mohd Arshad$^{2}$}
	\author{Ayyub Sheikhi$^{3}$
		\\\\ 
		$^{1}$D\lowercase{epartment of} M\lowercase{athematics and} S\lowercase{tatistics}, C\lowercase{entral} U\lowercase{niversity of} P\lowercase{unjab},
		B\lowercase{athinda}, P\lowercase{unjab}, I\lowercase{ndia}.
		%E\lowercase{mail: ashokiitb09@gmail.com}
		\\$^{2}$D\lowercase{epartment of} M\lowercase{athematics}, I\lowercase{ndian} I\lowercase{nstitute of} T\lowercase{echnology}
		I\lowercase{ndore}, S\lowercase{imrol}, I\lowercase{ndore}, I\lowercase{ndia}.
		\\$^{3}$D\lowercase{epartment} \lowercase{of} S\lowercase{tatistics}, S\lowercase{hahid} B\lowercase{ahonar} U\lowercase{niversity~of} K\lowercase{erman}, K\lowercase{erman}, I\lowercase{ran}.}
	%E\lowercase{mail: pv@math.iitb.ac.in}}
\thanks{Corresponding Author E-mail Address: ashokiitb09@gmail.com}%\\
%\thanks{The research of  Alam Ali was supported by DST Inspire Fellowship (IF190682), Government of India.}
% \thanks{E-mail address: ashokiitb09@gmail.com (Ashok Kumar Pathak).\\
	%The research of  A. K. Pathak was supported by RSM CUP Bathinda.
	%}
%\subjclass[2010]{Primary : 62E15, 62H05; Secondary : 62N05, 62H20}
%%%%%%%%%%%%%%%%%%%%%%%%%%%%%%%%%%%%%%%%%%%ABSTRACT%%%%%%%%%%%%%%%%%%%%%%%%%%%%%%%%%%%%%%%%%%%%%%%%%%%%%%%%%%
\vspace*{1in}
\begin{abstract}  Regression analysis is one of the most popularly used statistical technique which only measures the direct effect of independent variables on dependent variable. Path analysis looks for both direct and indirect effects of independent variables and may overcome  several hurdles allied with regression models. It utilizes one or more structural regression equations in the model which are used to estimate the unknown parameters. The aim of this work is to study the path analysis models when the endogenous (dependent) variable and exogenous (independent) variables are linked through the elliptical copulas.  Using well-organized numerical schemes,  we investigate the performance of path models when direct and indirect effects are estimated applying classical ordinary least squares  and copula-based regression approaches in different scenarios.  Finally, two real data applications are also presented to demonstrate the performance of path analysis  using copula approach. 
\end{abstract}

\maketitle
\section{Introduction} \noindent
Nowadays, researchers have a keen interest to analyze the hypothesized relationship among set of variables.
Path analysis is an extension of multiple regression, generally used to examine the importance of causal correlations among multiple variables. It was first introduced by Write \cite{wright1921correlation}. Basically, path analysis involves two components; path diagram and path coefficient which represents visual and mathematical parts of the analysis, respectively.
Primarily, path diagram analyzes causal relationship that occur in multiple regression when dependent variable is affected by independent variables directly and indirectly in both way (see Stage et al. \cite{stage2004path}). Smith et al. \cite{smith1997path} pointed out the ability of path analysis in a natural ecological system. G$\ddot{\text{u}}$neri et al. \cite{guneri2017path} investigated the distribution of indirect effect in path analysis. There are several applications of path analysis in field of sociology Ducan \cite{duncan1966path}, psychology Werts and Linn \cite{werts1970path}, epidemiology Karlin et al. \cite{karlin1983path}, and engineering  Huang and Hsueh \cite{huang2007study}.\\ 
\noindent The classical regression models are used to study the dependence between one dependent and one or more than one independent variables. It only measures the direct effect of independent variables on the dependent variable. In classical approach, path analysis is performed under the assumptions of classical regression (ordinary least squares (OLS)). Classical regression has certain limitations like the distribution of endogenous variable should be normal, relationship between endogenous and exogenous variables should be linear, and there should be no high multicollinearity occur among exogenous variables, but sometimes these assumptions are violated by the nature of data and strength of significant effect of one variable to another variable is not captured by the model. For more details, one may refer to Wiedermann et al. \cite{wiedermann2020direction}. Recently, copula-based regression models have been competent to handle the problem of non-normality and  multicollinearity in classical regression Kime and Kim \cite{kim2014analysis}, Kolev and Paiva \cite{kolev2009copula}, and Krupskii et al. \cite{krupskii2023factor}. Therefore, the copula-based regression models and its applications have been done by researchers in various field of statistics. Sungur \cite{sungur2005some} introduced the mathematical framework of copula-based regression model by considering uniform marginals. Crane and Hook \cite{crane2008conditional} derived the simple conditional expectation formulae in terms of copulas to carry out some regression analysis. Vellaisamy and Pathak \cite{vellaisamy2014copulas} introduced the concept of copula-based regression models for general random variables. In recent literature, the copula-based regression models have been also studied with an inferential aspect, see  Ali et al. \cite{alam2023}, Bouezmarni et al. \cite{bouezmarni2014regression}, Dette et al. \cite{dette2014some},  Kim et al. \cite{kim2007comparison}, Noh et al. \cite{noh2013copula}. Parsa and Klugman \cite{parsa2011copula} briefly discussed the advantages of copula models over ordinary least squares and generalized linear models. In literature, elliptical family of copulas, which are derived through multivariate elliptical distributions, consists of Gaussian and Student's t copula, are played vital role in dependence modeling. Using elliptical copulas, Frahm et al. \cite{frahm2003elliptical} studied the dependence structure induced via elliptical distributions. Regression model based on elliptical family of copulas is an useful probability model for the correlated data. Ghahroodi et al. \cite{ghahroodi2019gaussian} proposed a regression model using Gaussian copula to analyze the association among mixed outcomes. Pitt et al. \cite{pitt2006efficient} introduced a regression function using Gaussian copula and presented a general Bayesian approach to handle multivariate dependence. Acar et al. \cite{acar2019predictive} studied elliptical family of copulas for model based predictions. He et al. \cite{he2019robust} proposed a robust feature screening procedure for elliptical copulas regression model. Furthermore, Sheikhi et al. \cite{sheikhi2022heteroscedasticity} investigated a heteroscedasticity diagnostic in regression analysis when response and explanatory variables are connected through elliptical copulas. \\
Despite of availability of explored literature on copula-based regression models and path analysis, we observed that no one has performed the path analysis in the context of copula-based regression approach. Considering manifest (directly observed) variables, the primary aim of this article is to study the path analysis models when the endogenous (dependent) variable and exogenous (independent) variables are linked through the elliptical copulas and  to investigate the efficacy of path models when direct and indirect effects are estimated using classical ordinary least squares  and copula-based regression approaches in different scenarios. In this context, Braeken et al. \cite{braeken2013contextualized} have investigated the role of copulas in path analysis for categorical data. Douma and Shipley \cite{douma2023testing} have explored testing path models that include dependent errors, nonlinear functional relationships and using non-normal, hierarchically structured data. Choi and Seo \cite{choi2022copula} have proposed copula-based redundancy analysis to improve the performance of regression-based Extended Redundancy Analysis. Hult et al. \cite{hult2018addressing} have studied the effect of Gaussian Copula in Latent variables path analysis.  For more details, see Bauer \cite{bauer2013pair}, Eckert and Hohberger \cite{eckert2023addressing}, Park and Gupta \cite{park2012handling}.\\
In this work, based on a well organized empirical study and  real application, we found that path analysis performed through copula-based regression approach gives more robust results. The rest of work is organized as follows: Section 2 presents the  basic setup of regression based on copulas and path coefficients along with some important examples.  Section 3 summarizes the performance of considered approaches through well organized numerical scheme. To demonstrate the  performance of path analysis  through copula, two real data sets are also analyzed in Section 4. Finally, conclusion and some future remarks are discussed.
\section{Basic setup of Regression Based on Copulas and Path Coefficients}
\noindent Let $Y$ be a continuous endogenous variable and ${\bf X}=(X_1, X_2,\ldots, X_p)$ be continuous exogenous  random vector with joint distribution $H(y,\bf x)$ and  $H_{Y}(y)$ and $ H_{X_i}(x_i);i=1,2,\cdots p$,  be their marginal distributions, respectively. Then, by Sklar \cite{sklar1959fonctions} theorem, the marginals and joint distribution are connected via a $({p+1})$-dimensional copula $C$ as
\begin{equation*}
	H(y,x_1,\ldots,x_p)=C(H_{Y}(y), H_{X_1}(x_1),\ldots,H_{X_p}(x_p)),\;\;\forall\; (y,x_1,\ldots,x_p)\in\mathbb{R}^{p+1}.
\end{equation*} 
Let $H_{Y|\bf X}(y|{\bf x})$ be the conditional distribution of  $(Y,{\bf X})$,  then for a Borel measurable function $\eta$, the conditional expectation of $\eta(Y)$ on ${\bf X}={\bf x}$ is given by
\begin{equation*}
	m({\bf x})=E(\eta(Y)|{\bf X=\bf x})=\int \eta(y)\frac{\partial}{\partial y}H_{Y|\bf X}(y|{\bf x})dy.
\end{equation*}
In terms of copula, it is expressed as
\begin{equation}\label{M4cop1}
	m({\bf x})=E(\eta(Y)|{\bf X=\bf x})=\int \eta(y)\frac{c(H_{Y}(y), {H_{X_1}(x_1),\ldots,H_{X_p}(x_p)})}{c_{X}(H_{X_1}(x_1),\ldots,H_{X_p}(x_p))}dH_{Y}(y),
\end{equation}
where $\displaystyle c(u_0,u_1,\ldots,u_p)= \frac{\partial^{p+1}C(u_0,u_1,\ldots,u_p)}{\partial u_0\partial u_1\cdots \partial u_p}$ is the density of the copula $C$ associated with the random vector $(Y,{\bf X})$ and $\displaystyle c_{X}(u_1,\ldots,u_p)= \frac{\partial^{p}C(1,u_1,\ldots,u_p)}{\partial u_1\cdots \partial u_p}$ is the density of the copula associated with the random vector ${\bf X}$. When components of ${\bf X}$ are pairwise independent, then using fact that $\displaystyle c_{X}(u_1,\ldots,u_p)$, (\ref{M4cop1}) reduces to
\begin{equation}
	m({\bf x})=E(\eta(Y)|{\bf X=\bf x})=\int \eta(y){c(H_{Y}(y), H_{X_1}(x_1),\ldots,H_{X_p}(x_p)}dH_{Y}(y).
\end{equation}
Using (\ref{M4cop1}), we can derive higher order conditional moments, conditional variance and the regression function for different family of copulas. In recent literature, a number of researchers have studied the mathematical and statistical aspect of copula-based regression models. Some important references includes Ali et al. \cite{alam2023}, Crane and Hoek \cite{crane2008conditional}, Noh et al. \cite{noh2013copula}, Sheikhi et al. \cite{sheikhi2022heteroscedasticity}, and Vellaisamy and Pathak \cite{vellaisamy2014copulas}.\\  	
\noindent\textbf{{Regression for Elliptical Copulas Family}:}  Elliptical family of copulas are obtained  from the multivariate elliptical distribution functions.  It includes the Gaussian and the student's t-copula as two important members that are widely used in practical applications.  
Let $\boldsymbol{\vartheta}_{\Sigma_{p+1}}$ be the $(p+1)$-dimensional multivariate elliptical distribution with correlation matrix $\Sigma_{p+1}$ and $\vartheta^{-1}$ be the inverse distribution of the univariate elliptical distribution. Then, elliptical copulas are defined by 
\begin{equation}\label{M4eq2.1}
	C(u_0, u_1\ldots u_p)={ \boldsymbol{\vartheta}}_{\Sigma_{p+1}}\left(\vartheta^{-1}(u_0), \vartheta^{-1}(u_1),\ldots, \vartheta^{-1}(u_p)\right).
\end{equation}

\noindent If the joint distribution of $(Y, {\bf X})$ is determined by the elliptical copulas (\ref{M4eq2.1}), then regression of $Y$ on ${\bf X=\bf x}$ is 
\begin{equation*}
	m({\bf x})=\int y \frac{{ {\boldsymbol \nu}}_{\Sigma_{p+1}} \left(\vartheta^{-1}(H_Y(y),\vartheta^{-1}({{H_{X_1}(x_1)}},\ldots \vartheta^{-1}({{H_{X_p}(x_p)}})\right)}{{\boldsymbol\nu}_{\Sigma_{\bf X}} \left(\vartheta^{-1}({{H_{X_1}(x_1)}}, \ldots \vartheta^{-1}({{H_{X_p}(x_p)}})\right)}dy,
\end{equation*}
where ${ \boldsymbol\nu}_{\Sigma_{p+1}}$ and ${ \boldsymbol\nu}_{\Sigma_X}$ are joint densities of the multivariate  elliptical distributions of the random vectors $(Y,{\bf X})$ and ${\bf X}$ with correlation matrices ${\Sigma_{p+1}}$ and ${\Sigma_X}$, respectively. Next, we have the following results:
\begin{lemma}
	Let $C(u_0, u_1\ldots u_p)$ be a $(p+1)$-dimensional Gaussian copula of the form 
	\begin{equation*}
		C(u_0, u_1\ldots u_p)={\boldsymbol{\varPhi}}_{\Sigma_{p+1}}\left(\varPhi^{-1}(u_0),\varPhi^{-1}(u_1)\ldots \varPhi^{-1}(u_p) \right),
	\end{equation*}
	where $\boldsymbol{\varPhi}_{\Sigma_{p+1}}$ denotes a $(p+1)$-dimensional distribution of a standard normal variates, ${\varPhi^{-1}}$ represents the inverse of univariate standard normal. 
	Then, the regression of $Y$ on ${\bf X=\bf x}$ takes the form (see Ali et al. \cite{alam2023}, Noh et al. \cite{noh2013copula}). 
	\begin{equation}\label{M4N1}
		m({\bf x})=E\left[H^{-1}_{Y}\left(\varPhi\left({\bf u}^{'}\Sigma^{-1}_{X}\boldsymbol{\rho}+Z\sqrt{1-\boldsymbol{\rho}^{'}\Sigma^{-1}_{X}\boldsymbol{\rho}}\right)\right)\right],
	\end{equation}
	where $\boldsymbol{\rho}^{'}=\left(\rho_{x_1y},\ldots, \rho_{x_py}\right)$ presents the vector of correlation coefficient between $Y$ and ${\bf X}$ and ${\bf u}^{'}=\left(\varPhi^{-1}(H_{X_1}(x_1)),\ldots,\varPhi^{-1}(H_{X_p}(x_p))\right)$ denotes the vector of inverse cumulative distribution function of exogenous variables.
\end{lemma}
\begin{corollary} Let $Y, X_1, X_2 \sim N(0,1)$ and the joint dependence is determined by the Gaussian copula. 
	% Let $Y$ be the standard normal endogenous variable and $X_1, X_2$ be standard normal exogenous variables and joint dependence is determined by the Gaussian copula. 
	Then, the regression of $Y$ on $X_1=x_1$ and $X_2=x_2$ is given by  (see Sheikhi et al. \cite{sheikhi2022heteroscedasticity})
	\begin{equation}\label{M4G11}
		m(x_1,x_2)= \frac{\rho_{x_1y}-\rho_{x_2y}\rho_{x_1x_2}}{1-\rho^2_{x_1x_2}} x_1 +\frac{\rho_{x_2y}-\rho_{x_1y}\rho_{x_1x_2}}{1-\rho^2_{x_1x_2}} x_2,
	\end{equation}	
	where $\rho_{x_1y}, \rho_{x_2y}$, and $\rho_{x_1x_2}$ are correlation between $(X_1, Y)$, $(X_2, Y)$, and $(X_1, X_2)$, respectively.
	In terms of the path coefficients (\ref{M4G11}) may also be expressed as
	%Alternatively, (\ref{M4G11}) may be represented in terms of path coefficients as 
	\begin{equation*}
		m(x_1,x_2)=P_1 x_1 +P_2 x_2,
	\end{equation*}
	where $P_1=\displaystyle\frac{\rho_{x_1y}-\rho_{x_2y}\rho_{x_1x_2}}{1-\rho^2_{x_1x_2}}$ and $P_2=\displaystyle\frac{\rho_{x_2y}-\rho_{x_1y}\rho_{x_1x_2}}{1-\rho^2_{x_1x_2}}$ are path coefficients.
\end{corollary}
\begin{corollary} Let the joint dependence of ($Y$, $X_1$, $X_2$, $X_3$)  be characterized by Gaussian copula with standard normal marginals  
	%Let the endogenous variable ($Y$) and exogenous variables ($X_1$, $X_2$, and $X_3$)  are linked via Gaussian copula with standard normal marginals 
	and correlation coefficients among $Y$, $X_1$, $X_2$, and $X_3$  are $\rho_{x_1y}$, $\rho_{x_2y}$, $\rho_{x_3y}$, $\rho_{x_1x_2}$, $\rho_{x_1x_3}$, and $\rho_{x_2x_3}$, respectively. Then, the regression function of $Y$ on $X_1=x_1$,  $X_2=x_2 $, and $X_3=x_3 $ can be derived from \eqref{M4N1} and takes the following form   
	\begin{equation}\label{M4P1}
		\begin{aligned}
			m(x_1,x_2,x_3)&=
			\left\{\frac{\rho_{x_1y} (1-\rho^2_{x_2 x_3}) +\rho_{x_2y}  (\rho_{x_2x_3}\rho_{x_3x_1}-\rho_{x_2 x_1})+\rho_{x_3y}(\rho_{x_2x_1}\rho_{x_3x_2}-\rho_{x_3x_1})}{(1-\rho^2_{x_1x_2}-\rho^2_{x_1x_3}-\rho^2_{x_2x_3}+2\rho_{x_1x_2}\rho_{x_1x_3}\rho_{x_2x_3})}\right\}x_1\\
			&+ \left\{\frac{\rho_{x_2y} (1-\rho^2_{x_1 x_3}) +\rho_{x_1y} (\rho_{x_1x_3}\rho_{x_3x_2}-\rho_{x_1 x_2})+\rho_{x_3y}(\rho_{x_1x_2}\rho_{x_3x_1}-\rho_{x_3x_2})}{(1-\rho^2_{x_1x_2}-\rho^2_{x_1x_3}-\rho^2_{x_2x_3}+2\rho_{x_1x_2}\rho_{x_1x_3}\rho_{x_2x_3})}\right\}x_2\\
			&+ \left\{\frac{\rho_{x_3y} (1-\rho^2_{x_1 x_2}) +\rho_{x_1y} (\rho_{x_1x_2}\rho_{x_2x_3}-\rho_{x_1x_3})+\rho_{x_2y}(\rho_{x_1x_3}\rho_{x_2x_1}-\rho_{x_2x_3})}
			{(1-\rho^2_{x_1x_2}-\rho^2_{x_1x_3}-\rho^2_{x_2x_3}+2\rho_{x_1x_2}\rho_{x_1x_3}\rho_{x_2x_3})}\right\}x_3.
		\end{aligned}
	\end{equation}
	In terms of path coefficients ($P_j: j=1,2,3$) \eqref{M4P1} is written as
	\begin{equation*}
		m(x_1,x_2,x_3)=P_1 x_1 +P_2 x_2 + P_3 x_3,
	\end{equation*}
	where $P_1$, $P_2$, and $P_3$ are coefficients of $x_1$, $x_2$, and $x_3$ in \eqref{M4P1}.
\end{corollary}

%        \begin{align*}    	
	%        P_1&=\{\frac{\rho_{x_1y} (1-\rho^2_{x_2 x_3}) +\rho_{x_2y}  (\rho_{x_2x_3}\rho_{x_3x_1}-\rho_{x_2 x_1})+\rho_{x_3y}(\rho_{x_2x_1}\rho_{x_3x_2}-\rho_{x_3x_1})}{1-\rho^2_{x_1x_2}-\rho^2_{x_1x_3}-\rho^2_{x_2x_3}+2\rho_{x_1x_2}\rho_{x_1x_3}\rho_{x_2x_3}}\},\\  
	%        P_2&=\{\frac{\rho_{x_2y} (1-\rho^2_{x_1 x_3}) +\rho_{x_1y} (\rho_{x_1x_3}\rho_{x_3x_2}-\rho_{x_1 x_2})+\rho_{x_3y}(\rho_{x_1x_2}\rho_{x_3x_1}-\rho_{x_3x_2})}{1-\rho^2_{x_1x_2}-\rho^2_{x_1x_3}-\rho^2_{x_2x_3}+2\rho_{x_1x_2}\rho_{x_1x_3}\rho_{x_2x_3}}\}, \text{and}\\
	%        P_3&=\{\frac{\rho_{x_3y} (1-\rho^2_{x_1 x_2}) +\rho_{x_1y} (\rho_{x_1x_2}\rho_{x_2x_3}-\rho_{x_1x_3})+\rho_{x_2y}(\rho_{x_1x_3}\rho_{x_2x_1}-\rho_{x_2x_3})}
	%        {1-\rho^2_{x_1x_2}-\rho^2_{x_1x_3}-\rho^2_{x_2x_3}+2\rho_{x_1x_2}\rho_{x_1x_3}\rho_{x_2x_3}}\}.
	%        \end{align*}  
%		Further, in case of two exogenous variable, regression function takes the form (see \cite{sheikhi2022heteroscedasticityheteroscedasticity})
%		\begin{equation}
	%			\begin{aligned}\label{M4G11}
		%			m(x_1,x_2)&= \frac{\rho_{x_1y}-\rho_{x_2y}\rho_{x_1x_2}}{1-\rho^2_{x_1x_2}} x_1 +\frac{\rho_{x_2y}-\rho_{x_1y}\rho_{x_1x_2}}{1-\rho^2_{x_1x_2}} x_2,\\
		%			m(x_1,x_2)&=P_1 x_1 +P_2 x_2.
		%		\end{aligned}
	%		\end{equation}	  	  

%		For a special situation, when exogenous variables are pairwise independent i.e. $\rho_{x_1x_2}=0$, in tri-variate case regression function leads to
%		\begin{equation*} 
	%			m(x_1,x_2)= \rho_{x_1y} x_1 + \rho_{x_2y} x_2.
	%		\end{equation*}
\begin{lemma}
	Let $\Sigma_{p+1}$ be a $(p+1)$-dimensional correlation matrix.
	A Student's $t$-copula takes the form
	\begin{equation*}
		C(u_0, u_1\ldots u_p)={\bf t}_{\nu, \Sigma_{p+1}}\left(t^{-1}_{\nu}(u_0),t^{-1}_{\nu}({ u_1}),t^{-1}_{\nu}({ u_2})...t^{-1}_{\nu}({ u_p})\right),
	\end{equation*}
	where ${\bf t}_{\nu, \Sigma_{p+1}}$ is the multivariate Student's $t$-distribution with degrees of freedom $\nu$ and correlation matrix $\Sigma_{p+1}$, and $t^{-1}_{\nu}$ is  the inverse distribution function of univariate Student's $t$-distribution with $\nu$ degrees of freedom. Then, the conditional mean function of $Y$ on ${\bf X=x}$ is (see Leong and Valdez (\cite{leong2005claims})) given by
	\begin{equation*}\label{M4copregt}
		m({\bf x})=E\left[H^{-1}_{Y}\left(t_{\nu}(\boldsymbol{\rho}^{'}\Sigma^{-1}_{X}\boldsymbol{u}+\sqrt{\nu(1-\boldsymbol{\rho}^{'}\Sigma^{-1}_{X}\boldsymbol{\rho})\left(1+\frac{1}{\nu}\boldsymbol{u}^{'}\Sigma^{-1}_{X}\boldsymbol{u}\right)/(\nu+p)Z}\right)\right],
	\end{equation*}
	where ${\bf u}^{'}=\left(t^{-1}_{\nu}(H_{X_1}(x_1)),\ldots,t^{-1}_{\nu}(H_{X_p}(x_p))\right)$, $\boldsymbol{\rho}^{'}=\left(\rho_{x_1y},\ldots, \rho_{x_py}\right)$, $Z$ is standard univariate $t$ random variable.
\end{lemma}
\begin{remark}
	In case of real data applications, if the endogenous variable ($Y$) and exogenous variables ($X_1$ and $X_2$) are linked via t-copula with standard normal marginals and correlation coefficients among $Y$, $X_1$, and $X_2$  are $\rho_{x_1y}$, $\rho_{x_2y}$, and $\rho_{x_1x_2}$, respectively. Then, we can derive the algebraic expression of regression function discussed by Sheikhi et al. \cite{sheikhi2022heteroscedasticity} and for a large sample, this regression function tends to \eqref{M4G11}. 
\end{remark}
\begin{corollary} Let the joint dependence of ($Y$, $X_1$, $X_2$)
	%Let the joint distribution of endogenous variable ($Y$) and exogenous variables ($X_1$ and $X_2$) 
	is determined by the Student's t-copula with $\nu$ degree of freedom and all marginals are identical and have Student's $t$-distribution with location parameter ($\mu$) and scale parameter ($\sigma$). Then, in case of common pairwise correlation coefficient ($\rho$) between variables, regression function of  $Y$ on $X_1=x_1$ and $X_2=x_2$ takes the form (see Leong and Valdez \cite{leong2005claims})
	\begin{align*}
		m(x_1,x_2)&= \frac{1-\rho}{1+\rho } \mu+\frac{\rho}{1+\rho } x_1 + \frac{\rho}{1+\rho }x_2,\\
		m(x_1,x_2)&= \frac{1-\rho}{1+\rho } \mu + P_1 x_1 +P_2 x_2.
	\end{align*}
\end{corollary}	
\noindent With the help of above results, we can compute the path coefficient using copula function which measures the causal correlations between the endogenous and exogenous variables. Since, along with the direct effect of an exogenous variable, it may also have an indirect effect on an endogenous or another exogenous variable.  The correlation coefficient between these two variables is equal to the sum of the direct effect of an effective variable and indirect effects of the other variables (see G$\ddot{\text{u}}$neri et al. \cite{guneri2017path}). Therefore, the next result presents these correlations in terms of direct and indirect effects.
%Therefore, the mathematical representation of these correlations in terms of direct and indirect effects are given by the proposition (\ref{M4P11}). 	
\begin{proposition}{\label{M4P11}} 
	%	Let $Y$ be the continuous endogenous variable and $X_1, X_2,..., X_p $ be the continuous exogenous variables.
	Let the joint distribution of $(Y, X_1, X_2,..., X_p)$ is determined by elliptical distribution as defined in (\ref{M4eq2.1}). Let $Y$ and $X_i$ $(i=1,2,...,p)$ follow $N(0,1)$. Then, we can express the correlation between  endogenous and exogenous variables in terms of direct and indirect effects  as
	%	(see Guneri et al. \citeyear{guneri2017pathpath})
	\begin{align*}
		\rho_{x_1 x_1}P_1 + \rho_{x_1 x_2}P_2 +\cdots + \rho_{x_1 x_p}P_p &= \rho_{x_1y}\\
		\rho_{x_2 x_1}P_1 + \rho_{x_2 x_2}P_2 +\cdots + \rho_{x_2 x_p}P_p &= \rho_{x_2y}\\
		\vdots\hspace{2.5cm} \\
		\rho_{x_p x_1}P_1 + \rho_{x_p x_2}P_2 +\cdots + \rho_{x_p x_p}P_p &= \rho_{x_py}
	\end{align*}
	where, $P_j: j=1,2,...,p$ represents the direct effect of the exogenous variable ($X_i$) on the endogenous variable ($Y$) and
	$\rho_{x_i x_j}P_j: i \neq j=1,2,...,p $ denotes the indirect effect of an exogenous variable ($X_i$) on the $j^{th}$ exogenous variable ($X_j$).
\end{proposition} 
\noindent \textbf{Proof:} Let $Y= P_1X_1+ P_2X_2 + \cdots+ P_pX_p$. Then, the correlation between $Y$ and $X_1$ is given by 
%and all the variables have connected via Gaussian copula with standard normal margins (i.e. $\mu_y=\mu_{x_i}=0$ and $\sigma_y=\sigma_{x_i}=1; \forall i=1,2,...,p$). Then, using expectation ($E$) property, we can compute the correlation between $Y$ and $X_1$ such as 
%(see Niles and Henry E \citeyear{niles1922correlation})
\begin{center}
	$E(YX_1)= E[(P_1X_1+ P_2X_2 + \cdots+ P_pX_p)X_1].$ 
\end{center}
That is,
\begin{equation}\label{M4222}
	\sigma_{x_1y}=P_1\sigma^{2}_{x_1} + P_2\sigma_{x_1x_2} + \cdots + P_p\sigma_{x_1x_p}.
\end{equation}
Dividing both sides of (\ref{M4222}) by $\sigma_{y}\sigma_{x_1}$  and on rearranging the terms, it leads to
\begin{align*}\label{M4eq2}
	\frac{\sigma_{x_1y}}{\sigma_{y}\sigma_{x_1}}&=P_1\frac{\sigma^{2}_{x_1}}{\sigma_{y}\sigma_{x_1}} + P_2\frac{\sigma_{x_1x_2}}{\sigma_{y}\sigma_{x_1}}\times\frac{\sigma_{x_2}}{\sigma_{x_2}} + \cdots + P_p \frac{\sigma_{x_1x_p}}{\sigma_{y}\sigma_{x_1}}\times\frac{\sigma_{x_p}}{\sigma_{x_p}},
\end{align*}\\
which expresses the correlation between $Y$ and $X_1$ in terms of direct and indirect effects as
\begin{center}
	$ \rho_{x_1y}= P_1 + \rho_{x_1 x_2}P_2 + \cdots + \rho_{x_1 x_p}P_p.$
\end{center}	  
On a similar way, we can also obtain the other correlation between the variables ($Y, X_2$), ($Y, X_3$),...,($Y$, $X_p$) in terms of direct and indirect effects.  
\section{Simulation study}
\begin{figure}[h]
	\centering
	\subfloat{\includegraphics[scale=0.59, angle=0]{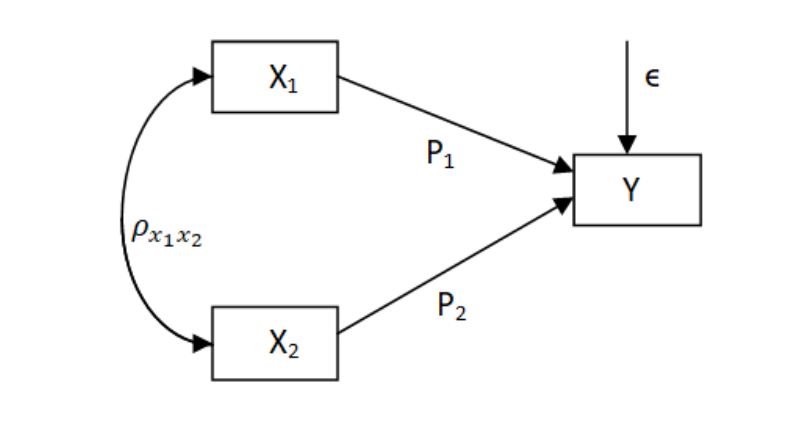}}
	\subfloat{\includegraphics[scale=0.55, angle=0]{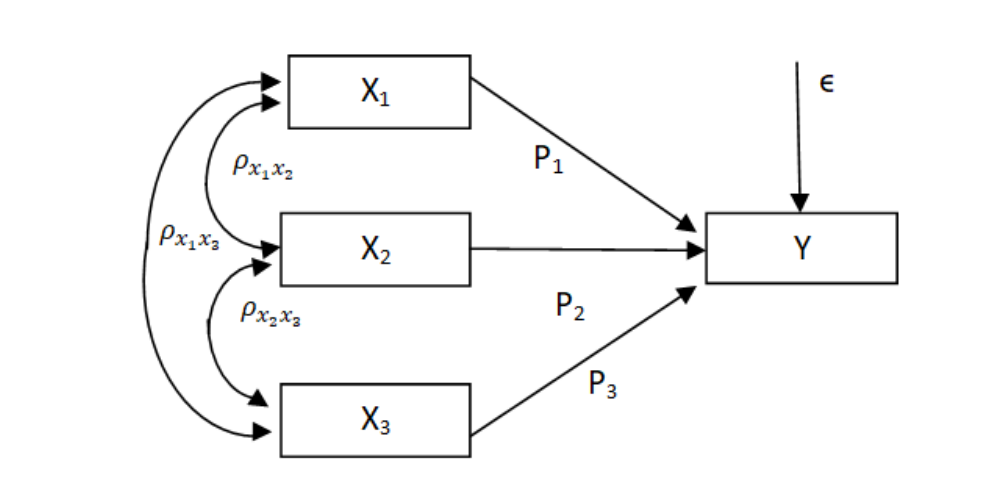}}
	\caption{ Path model in case of two exogenous variables (left) and three exogenous variables (right).}
	\label{M4Fig.1}
\end{figure}

\noindent 	Simulation study is carried out to estimate the causal effects in path analysis when variables are linked via Gaussian copula. This simulation also assesses the accuracy of path model when direct and indirect effect are estimated using classical and copula-based regression approaches. In this study, path model involves only observed variables. To deeply observe the performance of both approaches, we restrict our simulation on a tri-variate (which includes two exogenous variables $X_1$ and $X_2$, and one endogenous variable $Y$) and four-variate (three exogenous variables $X_1$, $X_2$, and $X_3$, and one endogenous variable $Y$)  path model. In both selected models, we define three different correlation levels; low, medium, and high among the variables. The path diagram of formulated causal models and the correlation structure of considered variables are shown in Fig \ref{M4Fig.1}. In the path model, the single-headed arrow shows the cause for one variable to another variable, while the double-headed arrow shows the correlation between pairs of exogenous variables. All the computations has been done through 4.2.1 version of R-programming.  We start our simulation using tri-variate  standard normal ($Y$, $X_1$, $X_2$) in which they are connected via Gaussian copula. 

% 	\clearpage
% \newpage
$\underbrace{
	\begin{array}{c}
		\begin{matrix}
			y &\hspace{0.2cm} x_1 &\hspace{0.2cm} x_2
		\end{matrix} \\
		\left[\ \begin{matrix}
			1 & 0.3 & -0.5 \\
			0.3 & 1 & 0.1 \\
			-0.5 & 0.1 & 1 
		\end{matrix}\ \right]
\end{array}}_{p=2,\hspace{0.1cm} Low\hspace{0.1cm} Correlation}$
\hspace{1.5cm}		
$\underbrace{
	\begin{array}{c}
		\begin{matrix}
			y &\hspace{0.2cm} x_1 &\hspace{0.2cm} x_2
		\end{matrix} \\
		\left[\ \begin{matrix}
			1 & 0.5 & 0.4 \\
			0.5 & 1 & 0.1 \\
			0.4  & 0.1 & 1
		\end{matrix}\ \right]
\end{array}}_{p=2,\hspace{0.1cm} Medium\hspace{0.1cm} Correlation}$	
\hspace{1.5cm}
$\underbrace{
	\begin{array}{c}
		\begin{matrix}
			y &\hspace{0.2cm} x_1 &\hspace{0.2cm} x_2
		\end{matrix} \\
		\left[\ \begin{matrix}
			1 & 0.6 & 0.7\\ 
			0.6 & 1 & 0.5\\
			0.7 & 0.5 & 1 
		\end{matrix}\ \right]
\end{array}}_{p=2,\hspace{0.1cm} High\hspace{0.1cm} Correlation}$	
\vspace{0.8cm}

$\underbrace{
	\begin{array}{c}
		\begin{matrix}
			y &\hspace{0.4cm} x_1 &\hspace{0.5cm} x_2 &\hspace{0.6cm} x_3
		\end{matrix} \\
		\left[\ \begin{matrix}
			1 & 0.3 & 0.2&-0.2\\ 
			0.3 & 1 & -0.1&0.1\\
			0.2 & -0.1 & 1 &0.2 \\
			-0.2 & 0.1 & 0.2 & 1 
		\end{matrix}\ \right]
\end{array}}_{p=3,\hspace{0.1cm} Low\hspace{0.1cm} Correlation}$
\hspace{0.6cm}
$\underbrace{
	\begin{array}{c}
		\begin{matrix}
			y &\hspace{0.2cm} x_1 &\hspace{0.2cm} x_2 &\hspace{0.2cm} x_3
		\end{matrix} \\
		\left[\ \begin{matrix}
			1 & 0.5& 0.4 & 0.3\\ 
			0.5 & 1&  0.2 & 0.1\\  
			0.4 & 0.2& 1 & 0.2 \\
			0.3 & 0.1 & 0.2 & 1
		\end{matrix}\ \right]
\end{array}}_{p=3,\hspace{0.1cm} Medium\hspace{0.1cm} Correlation}$
\hspace{0.6cm}
$\underbrace{
	\begin{array}{c}
		\begin{matrix}
			y &\hspace{0.2cm} x_1 &\hspace{0.2cm} x_2 &\hspace{0.2cm} x_3
		\end{matrix} \\
		\left[\ \begin{matrix}
			1 & 0.7 & 0.6 & 0.6\\ 
			0.7 & 1 & 0.5 & 0.5\\
			0.6 & 0.5 & 1 & 0.4 \\
			0.6 & 0.5 & 0.4 & 1
		\end{matrix}\ \right]
\end{array}}_{p=3,\hspace{0.1cm} High\hspace{0.1cm} Correlation}$\\      

\noindent 
We generate data of 100, 200, 300, and 400 different sample sizes from Gaussian copula in which endogenous and exogenous variables are linked with low, medium, and high correlations as defined above. Algorithm to generate data via Gaussian copula is briefly explained in `VineCopula' package in the R-software. Before performing path analysis, we first standardized data in case of each generated sample and confirmed the normality of data by Kolmogorov Samirov test. This test is performed using package `dgof' in R-software. Since the data comes from a Gaussian copula, it is evident that data hold the normality assumptions. Therefore, authors did not report the p-values of the K-S test corresponding to each model. Here, path analysis includes only one structure of regression equation based on the model pattern that has been already discussed in  Section 2. Since we use the 5-fold cross validation technique, so in each fold i.e., $k= 1, 2, 3, 4, 5$ train set contains only 80\% observations and remaining 20\% observations are contained in test set. To compare the performance of both approaches, we consider the statistical indices;  mean of mean squared error (MSE), standard deviation (SD) of  mean squared error (MSE),  Akaike Information Criterion (AIC), and Bayesian Information Criterion (BIC) of regression structure equation of path model in this 5-fold cross-validation technique. The mathematical form of these indices are as follows (see Sheikhi et al. \cite{sheikhi2022heteroscedasticity});
$	MSE=\frac{\sum_{1}^{n}(\hat{Y_i}-Y_i)^2}{n},$ $AIC= n LL + 2k,$ and  $BIC= n LL + k \log(n)$, where $n$ is the sample size of data, $LL$ denotes the log-likelihood for the model, and $k$ represents the number of parameters in the model. \\
In a particular situation when the path model consists of only two exogenous variable, Table \ref{M4Table1} presents the summary of model evaluation indices for the train set and test set corresponding to each low, medium, and high level correlation for 100, 200, 300, and 400 sample sizes. From Table \ref{M4Table1}, it can be observed that in case of train set; mean of MSE, SD of MSE, AIC, and BIC all are highly close in both approaches. But when we observe the performance of all indices for test set, analysis through the copula-based regression approach gives significant differences in results, which show the superiority of the Gaussian copula-based regression technique over the classical regression.\\
Table \ref{M4Table2} presents the path and correlation coefficients estimated through both techniques corresponding to low, medium, and high level correlation. The main observation of this table is that in the case of each sample size, the path coefficients calculated through classical and copula techniques are approximately equal to each other, and correlation coefficients are precisely similar in both approaches. Another observation is that increasing the sample size slightly affects the path coefficients. 	
We also observe the analysis of both techniques in terms of direct, indirect, and total effects. Table \ref{M4Table3} presents the summary of direct, indirect, and total effect for regression structure $P_1X_1$+$P_2X_2$ to $Y$ of path model. Path coefficients in classical approach (in case of a large sample, n=400) for train set reveals that there is a direct influence of $X_1$ on $Y$ (i.e., $P_1$) and $X_2$ on $Y$ (i.e., $P_2$) is 0.340 and -0.559, respectively. In other words, when the influence of one exogenous variable is kept constant, one-unit change in other exogenous variable in model will cause the relevant endogenous variable to change as much as the path coefficient value of the related variables. The results of  analysis show that indirect influence of  $X_1$ on $Y$ via $X_2$ (i.e., ${\rho_{x_1 x_2}}P_2$) and  $X_2$ on $Y$ via $X_1$ (i.e., ${\rho_{x_1 x_2}}P_1$)  are calculated as $0.103\times-0.559=-0.058$ and $0.103\times0.340 =0.035$, respectively, where 0.103 is the estimated value of correlation coefficient ($\rho_{x_1 x_2}$) between exogenous variables $X_1$ and $X_2$. The total effect of $X_1$ to $Y$ is $0.340+(-0.058)=0.282$ and $X_2$ to $Y$ is $(-0.599)+0.035=-0.523$, which are the approximately equal to the correlation coefficients between $X_1$ to $Y$ i.e., 0.3 and $X_2$ to $Y$ i.e., -0.5, respectively. Similarly, results of analysis obtained through Gaussian copula approach reveals the total effect of $X_1$ on $Y$ is 0.282  and $X_2$ on $Y$ is -0.523. 
Same way, when we interpretate the results for test set, the total effect of each exogenous variable $X_1$, $X_2$ on endogenous variable $Y$ are approximately equal to their correlation coefficient between $X_1$ on $Y$, and $X_2$ on $Y$, respectively (see proposition (\ref{M4P11})). 
%	We also observe a similar interpretation for other sample sizes. 
%	but in the case of large sample size, the total effect of each exogenous variable on the endogenous variable is approximately close to the correlation coefficient between these two variables as compared to the small sample sizes. 
Moreover, in case of large sample size, t-copula converges to the Gaussian copula (see Kole et al. \cite{kole2007selecting}) and regression function associated with Student's t-copula can be treated as a linear regression. For large samples, result of  analysis performed through t-copula is similar to the Gaussian copula. Therefore, these results are not reported in this manuscript.

\noindent Similarly, we use the same algorithm to perform  analysis when the path model consists of three exogenous and one endogenous variable.  From Table \ref{M4Table4}, Table \ref{M4Table5}, Table \ref{M4Table6}, and Table \ref{M4Table7}, we can observe the similar pattern of results as we obtained in the case of two exogenous and one endogenous path model.	
\begin{table}[h]
	\caption {Summary of model evaluation indices obtained by low, medium, and high correlations for $p=2$.}
	\centering
	%	\footnotesize
	\scriptsize
	%	\tiny
	\begin{tabular}{cccccccccc} % centered columns (6 columns)
		\toprule 
		Sample $\rightarrow$& \multicolumn{3}{c}{\hspace{2cm}$n = 100$} &\multicolumn{2}{c}{$n = 200$}& \multicolumn{2}{c}{$n = 300$} &\multicolumn{2}{c}{$n = 400$} \\
		\cmidrule{3-4} \cmidrule{5-6} \cmidrule{7-8} \cmidrule{9-10}	
		%		\hline
		%		
		%		Sample $\rightarrow$& &n=100&&n=200&&n=300&&n=400&\\
		$\downarrow$ Correlation &\multicolumn{1}{c}{Indices}&\multicolumn{1}{c}{Classical}& \multicolumn{1}{c}{Copula}&\multicolumn{1}{c}{Classical}& \multicolumn{1}{c}{Copula}&\multicolumn{1}{c}{Classical}& \multicolumn{1}{c}{Copula}&\multicolumn{1}{c}{Classical}& \multicolumn{1}{c}{Copula} \\ 
		\hline 
		Low & &&&&&&&&\\ 
		&Mean of MSE &0.539&0.540&0.585&0.587&0.556&0.556&0.609&0.609\\
		\parbox[t]{0.5cm}{\multirow{1}{*}{\rotatebox[origin=c]{90}{Train-Set}}}
		&SD of MSE  &0.013&0.013&0.032&0.031&0.022&0.022&0.012&0.011\\
		&	AIC  &-72.0&-71.5&-155.3&-155.6&-240.2&-239.5&-282.4&-281.9 \\
		&BIC  &-64.2&-63.6&-145.4&-144.5&-229.1&-228.5&-270.4&-269.8\\
		&  &&&&&&&&\\
		&Mean of MSE  &0.573&0.549&0.618&0.598&0.572&0.564&0.620&0.614\\
		\parbox[t]{0.5cm}{\multirow{1}{*}{\rotatebox[origin=c]{90}{Test-Set}}}
		&SD of MSE  &0.052&0.050&0.133&0.122&0.092&0.087&0.047&0.045\\
		&AIC  &-226.1&-231.8&-366.3&-378.9&-618.2&-627.7&-813.3&-820.8 \\
		&BIC  &-218.3&-223.9&-356.4&-369.1&-607.1&-616.6&-801.3&-808.9\\	
		\hline
		Medium & &&&&&&&&\\
		
		&Mean of MSE &0.570&0.570&0.574&0.574&0.634&0.634&0.636&0.636\\
		\parbox[t]{0.5cm}{\multirow{1}{*}{\rotatebox[origin=c]{90}{Train-Set}}}
		&SD of MSE  &0.039&0.038&0.019&0.019&0.016&0.016&0.021&0.021\\
		&	AIC  &-79.0&-78.9&-150.4&-150.3&-191.0&-190.9&-245.3&-245.1 \\
		&BIC  &-71.2&-71.1&-140.5&-140.4&-179.9&-179.8&-233.3&-233.1\\
		&  &&&&&&&&\\
		&Mean of MSE  &0.607&0.585&0.592&0.583&0.647&0.639&0.644&0.640\\
		\parbox[t]{0.5cm}{\multirow{1}{*}{\rotatebox[origin=c]{90}{Test-Set}}}
		&SD of MSE  &0.160&0.151&0.075&0.075&0.065&0.063&0.084&0.083\\
		&AIC  &-177.7&-178.0&-410.0&-411.1&-626.8&-628.0&-886.1&-890.1 \\
		&BIC  &-169.95&-169.98&-400.2&-401.2&-615.7&-617.1&-874.1&-878.1\\
		\hline
		High & &&&&&&&&\\
		
		&Mean of MSE &0.420&0.420&0.434&0.434&0.424&0.424&0.435&0.435\\
		\parbox[t]{0.5cm}{\multirow{1}{*}{\rotatebox[origin=c]{90}{Train-Set}}}
		&SD of MSE  &0.027&0.027&0.029&0.029&0.028&0.028&0.027&0.027\\
		&	AIC  &-107.8&-107.6&-208.1&-208.1&-327.6&-327.5&-437.3&-437.3 \\
		&BIC  &-100.0&-99.98&-198.2&-198.2&-316.5&-316.4&-425.3&-425.3\\
		&  &&&&&&&&\\
		&Mean of MSE  &0.473&0.453&0.456&0.442&0.432&0.428&0.439&0.436\\
		\parbox[t]{0.5cm}{\multirow{1}{*}{\rotatebox[origin=c]{90}{Test-Set}}}
		&SD of MSE  &0.119&0.117&0.126&0.118&0.112&0.110&0.109&0.108\\
		&AIC  &-209.9&-211.4&-458.2&-459.2&-685.0&-686.1&-884.3&-885.2 \\
		&BIC  &-202.1&-203.6&-448.3&-449.3&-673.9&-674.3&-872.3&-873.1 \\
		\hline
	\end{tabular}\label{M4Table1}
	
	\vspace{0.3cm}
\end{table}

\newpage
\begin{table}[h]
	\caption {Path coefficients ($P_i); i=1,2$ and correlation coefficient ($\rho_{x_1 x_2}$) obtained by low, medium, and high correlations for $p=2$.} % title of Table
	\centering
	%	\footnotesize
	\scriptsize
	%	\tiny
	\begin{tabular}{cccccccccc} % centered columns (6 columns)
		\toprule 
		Sample $\rightarrow$& \multicolumn{3}{c}{\hspace{0.7cm}$n = 100$} &\multicolumn{2}{c}{$n = 200$}& \multicolumn{2}{c}{$n = 300$} &\multicolumn{2}{c}{$n = 400$} \\
		\cmidrule{3-4} \cmidrule{5-6} \cmidrule{7-8} \cmidrule{9-10}	
		%		\hline
		%	Sample $\rightarrow$	& &n=100&&n=200&&n=300&&n=400&\\
		$\downarrow$	Correlation &\multicolumn{1}{c}{Coeff.}&\multicolumn{1}{c}{Classical}& \multicolumn{1}{c}{Copula}&\multicolumn{1}{c}{Classical}& \multicolumn{1}{c}{Copula}&\multicolumn{1}{c}{Classical}& \multicolumn{1}{c}{Copula}&\multicolumn{1}{c}{Classical}& \multicolumn{1}{c}{Copula} \\ 
		\hline 
		Low & &&&&&&&&\\
		\parbox[t]{0.5cm}{\multirow{1}{*}{\rotatebox[origin=c]{90}{Train-Set}}} 
		&$P_1$ &0.3588&0.3578&0.3841&0.3839&0.3442&0.3438&0.3404&0.3402\\
		&$P_2$  &-0.6128&-0.6095&-0.5740&-0.5737&-0.6044&-0.6042&-0.5589&-0.5585\\
		&${\rho_{x_1 x_2}}$ &0.1036&0.1036&0.1478&0.1478&0.0992&0.0992&0.1030&0.1030\\
		&  &&&&&&&&\\
		\parbox[t]{0.5cm}{\multirow{1}{*}{\rotatebox[origin=c]{90}{Test-Set}}} 
		&$P_1$ &0.3588&0.3633&0.3841&0.4122&0.3442&0.3534&0.3404&0.3385\\
		&$P_2$  &-0.6128&-0.6493&-0.5740&-0.6016&-0.6044&-0.6145&-0.5589&-0.5638\\	
		&${\rho_{x_1 x_2}}$  &0.1081&0.1081&0.1564&0.1564&0.1041&0.1041&0.1044&0.1044\\
		\hline
		Medium & &&&&&&&&\\
		\parbox[t]{0.5cm}{\multirow{1}{*}{\rotatebox[origin=c]{90}{Train-Set}}} 
		&$P_1$ &0.5354&0.5353&0.5225&0.5227&0.4738&0.4739&0.4874&0.4873\\
		&$P_2$  &0.3241&0.3236&0.3553&0.3554&0.3632&0.3628&0.3562&0.3562\\
		&${\rho_{x_1 x_2}}$  &0.0835&0.0835&0.0586&0.0586&0.0199&0.0199&-0.0078&-0.0078\\
		&  &&&&&&&&\\
		\parbox[t]{0.5cm}{\multirow{1}{*}{\rotatebox[origin=c]{90}{Test-Set}}} 
		&$P_1$ &0.5354&0.5068&0.5225&0.5262&0.4738&0.4819&0.4874&0.4960\\
		&$P_2$  &0.3241&0.3585&0.3553&0.3639&0.3632&0.3565&0.3562&0.3508\\	
		&${\rho_{x_1 x_2}}$  &0.0905&0.0905&0.0540&0.0540&0.0165&0.0165&-0.0058&-0.0058\\
		\hline
		High & &&&&&&&&\\
		\parbox[t]{0.5cm}{\multirow{1}{*}{\rotatebox[origin=c]{90}{Train-Set}}} 
		&$P_1$ &0.3270&0.3288&0.3419&0.3424&0.3650&0.3653&0.3373&0.3374\\
		&$P_2$  &0.5398&0.5397&0.5206&0.5204&0.5145&0.5143&0.5272&0.5271\\
		&${\rho_{x_1 x_2}}$  &0.4727&0.4727&0.4868&0.4868&0.4665&0.4665&0.4821&0.4821 \\
		&  &&&&&&&&\\
		\parbox[t]{0.5cm}{\multirow{1}{*}{\rotatebox[origin=c]{90}{Test-Set}}} 
		&$P_1$ &0.3270&0.3931&0.3419&0.3562&0.3650&0.3660&0.3373&0.3350 \\
		&$P_2$  &0.5398&0.4849&0.5206&0.5125&0.5145&0.5190&0.5272&0.5294\\	
		&${\rho_{x_1 x_2}}$  &0.4616&0.4616&0.4783&0.4783&0.4640&0.4640&0.4759&0.4759\\
		\hline
	\end{tabular}\label{M4Table2}
\end{table} 
%\noindent Similarly, we use the same algorithm to perform  analysis when the path model consists of three exogenous and one endogenous variable.  From Table \ref{M4Table4}, Table \ref{M4Table5}, Table \ref{M4Table6}, and Table \ref{M4Table7}, we can observe the similar pattern of results as we obtained in the case of two exogenous and one endogenous path model.	

\newpage
\begin{table}[h!]
	\caption{Summary of direct effect (D.E.), indirect effect (I.E.), and total effect (T.E.) for $p=2$.} % title of Table
	\centering
	%	\footnotesize
	%	\scriptsize
	\tiny
	
	\begin{tabular}{cccccccccccccc} % centered columns (6 columns)
		\toprule 
		Sample $\rightarrow$& \multicolumn{3}{c}{\hspace{2.5cm}$n = 100$}& &\multicolumn{2}{c}{\hspace{0.7cm}$n = 200$}&& \multicolumn{2}{c}{\hspace{0.7cm}$n = 300$} &&\multicolumn{2}{c}{\hspace{0.7cm}$n = 400$}&\\
		\cmidrule{3-5} \cmidrule{6-8} \cmidrule{9-11} \cmidrule{12-14}
		$\downarrow$ Corr.&Approach&D.E. &I.E.&T.E.&D.E. &I.E.&T.E.&D.E. &I.E.&T.E.&D.E. &I.E.&T.E. \\ [1ex] % inserts table
		\hline
		Low&\multicolumn{1}{c}{Classical}& \multicolumn{1}{c}{} & \multicolumn{1}{c}{} & \multicolumn{1}{c}{}&&&&&&&&&\\ 	
		&$X_1$ to $Y$ &0.359&-0.064&0.295&0.384&-0.085&0.299&0.344&-0.060&0.284&0.340&-0.058&0.282 \\
		\parbox[t]{5mm}{\multirow{1}{*}{\rotatebox[origin=c]{90}{Train-Set }}}
		&$X_2$ to $Y$ &-0.613&0.038&-0.575&-0.574&0.057&-0.517&-0.604&0.034&-0.570&-0.559&0.035&-0.523\\
		&Copula & & &&&&&&&&&& \\
		&$X_1$ to $Y$ &0.358&-0.064&0.294&0.384&-0.085&0.299&0.344&-0.061&0.283&0.340&-0.058&0.282 \\
		&$X_2$ to $Y$ &-0.610&0.038&-0.572&-0.574&0.058&-0.516&-0.604&0.034&-0.570&-0.559&0.036&-0.523 \\
		
		%&\multicolumn{1}{c}{}& \multicolumn{1}{c}{} & \multicolumn{1}{c}{} & \multicolumn{1}{c}{}\\ 	
		Low&Classical &&&&&&&&&&&& \\
		&$X_1$ to $Y$&0.359&-0.067&0.292&0.384&-0.090&0.294&0.344&-0.063&0.281&0.340&-0.058&0.282 \\
		\parbox[t]{5mm}{\multirow{1}{*}{\rotatebox[origin=c]{90}{Test-Set }}}
		&$X_2$ to $Y$ &-0.613&0.039&-0.574&-0.574&0.061&-0.513&-0.604&0.036&-0.568&-0.559&0.076&-0.523 \\
		&Copula &&&&&&&&&&&& \\	
		&$X_1$ to $Y$ &0.363&-0.070&0.293&0.412&-0.094&0.318&0.353&-0.064&0.289&0.339&-0.06&0.279 \\
		&$X_2$ to $Y$ &-0.649&0.039&-0.610&-0.602&0.065&-0.537&-0.615&0.038&-0.577&-0.564&0.036&-0.528 \\
		\hline
		
		Medium&\multicolumn{1}{c}{Classical}& \multicolumn{1}{c}{} & \multicolumn{1}{c}{} & \multicolumn{1}{c}{}&&&&&&&&&\\ 		
		&$X_1$ to $Y$ &0.535&0.027&0.562&0.523&0.020&0.543&0.474&0.007&0.481&0.487&-0.002&0.485 \\
		\parbox[t]{5mm}{\multirow{1}{*}{\rotatebox[origin=c]{90}{Train-Set}}}
		&$X_2$ to $Y$ &0.324&0.044&0.368&0.355&0.031&0.386&0.363&0.010&0.373&0.356&-0.004&0.352\\
		&Copula & & &&&&&&&&&& \\
		&$X_1$ to $Y$ &0.535&0.027&0.562&0.523&0.021&0.544&0.474&0.007&0.481& 0.487&-0.003&0.484 \\
		&$X_2$ to $Y$ &0.324&0.004&0.368&0.355&0.031&0.386&0.363&0.009&0.372&0.356&-0.004& 0.352\\
		
		%&\multicolumn{1}{c}{}& \multicolumn{1}{c}{} & \multicolumn{1}{c}{} & \multicolumn{1}{c}{}\\ 	
		Medium&Classical &&&&&&&&&&&& \\
		&$X_1$ to $Y$&0.535&0.029&0.564&0.523&0.019&0.542&0.474&0.006&0.480&0.487&-0.002&0.485 \\
		\parbox[t]{5mm}{\multirow{1}{*}{\rotatebox[origin=c]{90}{Test-Set}}}
		&$X_2$ to $Y$ &0.324&0.048&0.372&0.355&0.029&0.384&0.363&0.008&0.371&0.356&-0.003&0.353\\
		&Copula &&&&&&&&&&&& \\	
		&$X_1$ to $Y$ &0.507&0.032&0.539&0.526&0.020&0.546&0.482&0.006&0.488&0.496&-0.002&0.494 \\
		&$X_2$ to $Y$ &0.359&0.045&0.404&0.364&0.028&0.392&0.357&0.007&0.364&0.351&-0.003&0.348 \\
		\hline
		
		High&\multicolumn{1}{c}{Classical}& \multicolumn{1}{c}{} & \multicolumn{1}{c}{} & \multicolumn{1}{c}{}&&&&&&&&&\\ 
		&$X_1$ to $Y$ &0.327&0.255&0.582&0.342&0.253&0.595&0.365&0.240&0.605&0.337&0.254&0.591 \\
		\parbox[t]{5mm}{\multirow{1}{*}{\rotatebox[origin=c]{90}{Train-Set }}}
		&$X_2$ to $Y$ &0.540&0.154&0.694&0.520&0.167&0.687&0.515&0.169&0.684&0.527&0.163&0.690 \\
		&Copula & & &&&&&&&&&& \\
		&$X_1$ to $Y$ &0.329&0.254&0.583&0.342&0.253&0.595&0.365&0.240&0.605&0.337&0.255&0.592 \\
		&$X_2$ to $Y$ &0.540&0.155&0.695&0.520&0.167&0.687&0.514&0.171&0.685&0.527&0.163&0.690 \\
		
		%&\multicolumn{1}{c}{}& \multicolumn{1}{c}{} & \multicolumn{1}{c}{} & \multicolumn{1}{c}{}\\ 	
		High&Classical &&&&&&&&&&&& \\
		&$X_1$ to $Y$&0.327&0.249&0.576&0.342&0.248&0.590&0.365&0.239&0.604&0.337&0.251&0.588 \\
		\parbox[t]{5mm}{\multirow{1}{*}{\rotatebox[origin=c]{90}{Test-Set}}}
		&$X_2$ to $Y$ &0.540&0.151&0.691&0.521&0.163&0.684&0.515&0.169&0.684&0.527&0.161&0.688 \\
		&Copula &&&&&&&&&&&& \\	
		&$X_1$ to $Y$ &0.393&0.223&0.616&0.356&0.245&0.601&0.366&0.241&0.607&0.335&0.252&0.587 \\
		&$X_2$ to $Y$ &0.485&0.181&0.666&0.513&0.170&0.683&0.519&0.170&0.689&0.529&0.160&0.689 \\
		\hline
	\end{tabular}\label{M4Table3}
\end{table}

\newpage
\begin{table}[h!]
	\caption {Summary of model evaluation indices obtained by low, medium, and high correlations for $p=3$. } % title of Table
	\centering
	%	\footnotesize
	\scriptsize
	%	\tiny
	\begin{tabular}{cccccccccc} % centered columns (6 columns)
		\toprule 
		Sample $\rightarrow$& \multicolumn{3}{c}{\hspace{1.6cm}$n = 100$} &\multicolumn{2}{c}{$n = 200$}& \multicolumn{2}{c}{$n = 300$} &\multicolumn{2}{c}{$n = 400$} \\
		\cmidrule{3-4} \cmidrule{5-6} \cmidrule{7-8} \cmidrule{9-10}	
		%		\hline
		%		& &n=100&&n=200&&n=300&&n=400&\\
		$\downarrow$ Correlation &\multicolumn{1}{c}{Indices}&\multicolumn{1}{c}{Classical}& \multicolumn{1}{c}{Copula}&\multicolumn{1}{c}{Classical}& \multicolumn{1}{c}{Copula}&\multicolumn{1}{c}{Classical}& \multicolumn{1}{c}{Copula}&\multicolumn{1}{c}{Classical}& \multicolumn{1}{c}{Copula} \\ 
		\hline 
		Low & &&&&&&&&\\ 
		&Mean of MSE &0.676&0.676&0.686&0.687&0.750&0.751&0.748&0.749\\
		\parbox[t]{0.5cm}{\multirow{1}{*}{\rotatebox[origin=c]{90}{Train-Set}}}
		&SD of MSE  &0.058&0.058&0.037&0.037&0.027&0.028&0.030&0.030\\
		&	AIC  &-59.9&-59.9&-120.2&-120.2&-154.3&-154.1&-211.6&-211.3 \\
		&BIC  &-44.4&-44.3&-100.4&-100.4&-132.1&-131.8&-187.7&-187.4\\
		&  &&&&&&&&\\
		&Mean of MSE  &0.740&0.616&0.737&0.644&0.790&0.718&0.789&0.706\\
		\parbox[t]{0.5cm}{\multirow{1}{*}{\rotatebox[origin=c]{90}{Test-Set}}}
		&SD of MSE  &0.246&0.215&0.145&0.140&0.107&0.126&0.119&0.135\\
		&AIC  &-130.5&-144.1&-314.5&-330.4&-480.4&-488.8&-650.0&-665.6 \\
		&BIC  &-114.9&-128.4&-294.7&-310.6&-458.1&-466.5&-626.1&-641.6\\	
		\hline
		Medium & &&&&&&&&\\
		
		&Mean of MSE &0.594&0.594&0.571&0.571&0.593&0.593&0.611&0.611\\
		\parbox[h!]{0.5cm}{\multirow{1}{*}{\rotatebox[origin=c]{90}{Train-Set}}}
		&SD of MSE  &0.034&0.034&0.017&0.017&0.018&0.018&0.020&0.020\\
		&	AIC  &-64.0&-63.9&-133.9&-133.5&-209.8&-209.7&-284.4&-283.8 \\
		&BIC  &-48.4&-48.4&-114.1&-113.8&-187.6&-187.5&-260.5&-259.9\\
		&  &&&&&&&&\\
		&Mean of MSE  &0.675&0.527&0.618&0.531&0.624&0.553&0.627&0.594\\
		\parbox[t]{0.5cm}{\multirow{1}{*}{\rotatebox[origin=c]{90}{Test-Set}}}
		&SD of MSE  &0.154&0.119&0.076&0.038&0.073&0.079&0.082&0.077\\
		&AIC  &-165.6&-176.3&-428.4&-443.2&-602.2&607.9&-759.5&-780.9 \\
		&BIC  &-149.9&-160.6&-408.6&-423.4&-580.0&-585.7&-735.5&-756.9\\
		\hline
		High & &&&&&&&&\\
		
		&Mean of MSE &0.314&0.315&0.401&0.402&0.404&0.404&0.395&0.395\\
		\parbox[h!]{0.5cm}{\multirow{1}{*}{\rotatebox[origin=c]{90}{Train-Set}}}
		&SD of MSE  &0.032&0.031&0.020&0.020&0.019&0.019&0.018&0.018\\
		&	AIC  &-110.9&-110.6&-199.6&-199.4&-312.6&-312.4&-428.1&-428.0 \\
		&BIC  &-95.1&-94.9&-179.8&-179.6&-290.4&-290.2&-404.2&-404.1\\
		&  &&&&&&&&\\
		&Mean of MSE  &0.360&0.287&0.419&0.388&0.417&0.392&0.402&0.389\\
		\parbox[t]{0.5cm}{\multirow{1}{*}{\rotatebox[origin=c]{90}{Test-Set}}}
		&SD of MSE  &0.145&0.087&0.083&0.078&0.077&0.072&0.074&0.071\\
		&AIC  &-296.5&-313.8&-526.5&-541.5&-767.8&-784.5&-1063.7&-1072.5 \\
		&BIC  &-280.9&-298.2&-506.7&-521.8&-745.6&-762.3&-1039.7&-1048.5\\
		\hline
	\end{tabular}\label{M4Table4}
\end{table}
\clearpage
\newpage
\begin{table}[h!]
	\caption {Path coefficients ($P_i);i=1,2,3$ and correlation coefficient (${\rho_{x_1 x_2}}$, ${\rho_{x_1 x_3}}$, ${\rho_{x_2 x_3}}$) obtained by low, medium, and high correlations for $p=3$.} % title of Table
	\centering
	%	\footnotesize
	%	\scriptsize
	\tiny
	\begin{tabular}{cccccccccc} % centered columns (6 columns)	
		\toprule 
		Sample $\rightarrow$& \multicolumn{3}{c}{\hspace{0.7cm}$n = 100$} &\multicolumn{2}{c}{$n = 200$}& \multicolumn{2}{c}{$n = 300$} &\multicolumn{2}{c}{$n = 400$} \\
		\cmidrule{3-4} \cmidrule{5-6} \cmidrule{7-8} \cmidrule{9-10}
		%	\hline
		%	Sample $\rightarrow$	& &n=100&&n=200&&n=300&&n=400&\\
		$\downarrow$	Corr. &\multicolumn{1}{c}{Coeff.}&\multicolumn{1}{c}{Classical}& \multicolumn{1}{c}{Copula}&\multicolumn{1}{c}{Classical}& \multicolumn{1}{c}{Copula}&\multicolumn{1}{c}{Classical}& \multicolumn{1}{c}{Copula}&\multicolumn{1}{c}{Classical}& \multicolumn{1}{c}{Copula} \\ 
		\hline 
		Low&$P_1$ &0.3770&0.3770&0.4124&0.4119&0.3763&0.3758&0.3846&0.3838\\	
		&$P_2$  &0.3605&0.3597&0.3171&0.3158&0.2875&0.2866&0.3111&0.3101\\
		\parbox[t]{0.5cm}{\multirow{1}{*}{\rotatebox[origin=c]{90}{Train-Set }}}
		&$P_3$  &-0.3334&-0.3326&-0.3085&-0.3085&-0.2901&-0.2897&-0.3000&-0.2993\\
		&${\rho_{x_1 x_2}}$ &0.0255&0.0255&-0.0391&-0.0391&-0.1101&-0.1101&-0.1423&-0.1423\\
		&${\rho_{x_1 x_3}}$ &0.2126&0.2126&0.1560&0.1560&0.1513&0.1513&0.1437&0.1437\\
		&${\rho_{x_2 x_3}}$ &0.1050&0.1050&0.0567&0.0567&0.0476&0.0476&0.1111&0.1111\\
		
		Low&$P_1$ &0.3770&0.3493&0.4124&0.3881&0.3763&0.3642&0.3846&0.3676\\
		&$P_2$  &0.3605&0.3626&0.3171&0.3259&0.2875&0.2837&0.3111&0.3123\\
		\parbox[t]{0.5cm}{\multirow{1}{*}{\rotatebox[origin=c]{90}{Test-Set }}} 	
		&$P_3$  &-0.3334&-0.3348&-0.3085&-0.2919&-0.2901&-0.2828&-0.3000&-0.2805\\	
		&${\rho_{x_1 x_2}}$  &0.0557&0.0557&-0.0161&-0.0161&-0.0976&-0.0976&-0.1224&-0.1224\\
		&${\rho_{x_1 x_3}}$ &0.2237&0.2237&0.1594&0.1594&0.1577&0.1577&0.1579&0.1579\\
		&${\rho_{x_2 x_3}}$ &0.1007&0.1007&0.0564&0.0564&0.0421&0.0421&0.1050&0.1050\\
		\hline
		Medium&$P_1$ &0.4497&0.4493&0.4462&0.4465&0.4154&0.4150&0.4145&0.4146\\	
		&$P_2$  &0.3015&0.2984&0.2850&0.2839&0.2919&0.2919&0.2863&0.2864\\
		\parbox[t]{0.5cm}{\multirow{1}{*}{\rotatebox[origin=c]{90}{Train-Set}}}
		&$P_3$  &0.1055&0.1062&0.2150&0.2156&0.2253&0.2246&0.2117&0.2110\\
		&${\rho_{x_1 x_2}}$ &0.2528&0.2528&0.1977&0.1977&0.1877&0.1877&0.1537&0.1537\\
		&${\rho_{x_1 x_3}}$ &0.1088&0.1088&0.0969&0.0969& 0.1143& 0.1143&0.1180&0.1180\\
		&${\rho_{x_2 x_3}}$ &0.1943&0.1943&0.1996&0.1996&0.2070&0.2070&0.2535&0.2535\\
		
		Medium&$P_1$ &0.4497&0.4434& 0.4462&0.4249&0.4154&0.4190&0.4145&0.4191\\ 
		&$P_2$  &0.3015&0.2942&0.2850&0.2692&0.2919&0.2774&0.2863&0.2822\\	
		\parbox[t]{0.5cm}{\multirow{1}{*}{\rotatebox[origin=c]{90}{Test-Set }}}
		&$P_3$  &0.1055&0.1343&0.2150&0.2238&0.2253&0.2315&0.2117&0.2142\\	
		&${\rho_{x_1 x_2}}$  &0.2312&0.2312&0.1721&0.1721&0.1854&0.1854&0.1484&0.1484\\
		&${\rho_{x_1 x_3}}$ &0.1196&0.1196&0.0968&0.0968&0.1159&0.1159&0.1173&0.1173\\
		&${\rho_{x_2 x_3}}$ &0.2095&0.2095&0.2147&0.2147&0.2137&0.2137&0.2546&0.2546\\
		\hline
		High&$P_1$ &0.6050&0.6054&0.4362&0.4359&0.3896&0.3891&0.4025&0.4020\\	
		&$P_2$  &0.2063&0.2046&0.2453&0.2453&0.2749&0.2749&0.2568&0.2569\\
		\parbox[t]{0.5cm}{\multirow{1}{*}{\rotatebox[origin=c]{90}{Train-Set }}}
		&$P_3$  &0.1404&0.1399&0.2719&0.2453&0.2910&0.2907&0.3032&0.3030\\
		&${\rho_{x_1 x_2}}$ &0.5470&0.5470&0.5225&0.5225&0.4970&0.4970&0.4713&0.4713\\
		&${\rho_{x_1 x_3}}$ &0.5470&0.5470&0.4837&0.4837&0.4905&0.4905&0.5031&0.5031\\
		&${\rho_{x_2 x_3}}$ &0.3920&0.3920&0.3356&0.3356&0.4057&0.4057&0.4152&0.4152\\
		
		High&$P_1$ &0.6050&0.6372&0.4362&0.4456&0.3896&0.3896&0.4025&0.3987\\
		&$P_2$  &0.2063&0.1564&0.2453&0.2234&0.2749&0.2710&0.2568&0.2546\\
		\parbox[t]{0.5cm}{\multirow{1}{*}{\rotatebox[origin=c]{90}{Test-Set }}} 	
		&$P_3$  &0.1404&0.1120&0.2719&0.2719&0.2910&0.2889&0.3032& 0.3068\\	
		&${\rho_{x_1 x_2}}$  &0.5113&0.5153&0.5395&0.5395&0.5166&0.5166&0.4862&0.4862\\
		&${\rho_{x_1 x_3}}$ &0.5812&0.5812&0.4856&0.4856&0.4945&0.4945&0.5070&0.5070\\
		&${\rho_{x_2 x_3}}$ &0.4179&0.4179&0.3205&0.3205&0.4131&0.4131&0.4260&0.4260\\
		\hline
		
	\end{tabular}\label{M4Table5}
\end{table}
\newpage
\begin{table}[h!]
	\caption{Summary of direct effect (D.E.), indirect effect (I.E.), and total effect (T.E.) obtained by low and medium  correlations for $p=3$.} % title of Table
	\centering
	%	\footnotesize
	%		\scriptsize
	\tiny
	
	\begin{tabular}{cccccccccccccc} % centered columns (6 columns)
		\toprule 
		Sample $\rightarrow$& \multicolumn{3}{c}{\hspace{2cm}$n = 100$}& &\multicolumn{2}{c}{\hspace{0.7cm}$n = 200$}&& \multicolumn{2}{c}{\hspace{0.7cm}$n = 300$} &&\multicolumn{2}{c}{\hspace{0.7cm}$n = 400$}&\\
		\cmidrule{3-5} \cmidrule{6-8} \cmidrule{9-11} \cmidrule{12-14}	
		%		\hline
		%		\hline
		%		Sample $\rightarrow$& &n=100&&&n=200&&&n=300&&&n=400&&\\
		$\downarrow$ Corr.&Approach&D.E. &I.E.&T.E.&D.E. &I.E.&T.E.&D.E. &I.E.&T.E.&D.E. &I.E.&T.E. \\ [1ex] % inserts table
		\hline
		Low&\multicolumn{1}{c}{Classical}& \multicolumn{1}{c}{} & \multicolumn{1}{c}{} & \multicolumn{1}{c}{}&&&&&&&&&\\ 	
		&$X_1$ to $Y$&0.377&-0.062&0.315&0.412&-0.060&0.352&0.376&-0.075&0.301&0.385&-0.008&0.297  \\
		
		&$X_2$ to $Y$&0.361&-0.026&0.335&0.317&-0.034&0.283&0.288&-0.056&0.232&0.311&-0.088&0.223 \\
		\parbox[t]{5mm}{\multirow{1}{*}{\rotatebox[origin=c]{90}{Train-Set }}}
		&$X_3$ to $Y$&-0.333&0.118&-0.215&-0.309&0.083&-0.226&-0.290&0.071&-0.219&-0.300&0.090&-0.210 \\
		
		&Copula & & &&&&&&&&&& \\
		&$X_1$ to $Y$&0.377&-0.062&0.315&0.412&-0.061&0.351&0.376&-0.076&0.300&0.384&-0.087&0.297 \\
		&$X_2$ to $Y$&0.360&-0.026&0.334&0.316&-0.034&0.282&0.287&-0.056&0.231&0.310&-0.088&0.222  \\
		&$X_3$ to $Y$&-0.333&0.118&-0.215&-0.309&0.083&-0.226&-0.290&0.071&-0.219&-0.299&0.089& -0.210\\
		
		%&\multicolumn{1}{c}{}& \multicolumn{1}{c}{} & \multicolumn{1}{c}{} & \multicolumn{1}{c}{}\\ 	
		Low&Classical &&&&&&&&&&&& \\
		&$X_1$ to $Y$&0.377&-0.055&0.322&0.412&-0.054&0.358&0.376&-0.074&0.302&0.385&-0.086&0.299 \\
		
		&$X_2$ to $Y$&0.361&-0.012&0.349&0.317&-0.024&0.293&0.288&-0.049&0.239&0.311&-0.078&0.233  \\
		\parbox[t]{5mm}{\multirow{1}{*}{\rotatebox[origin=c]{90}{Test-Set }}}
		&$X_3$ to $Y$&-0.333&0.120&-0.213&-0.309&0.084&-0.225&-0.290&0.071&-0.219&-0.300&0.093&-0.207 \\
		
		&Copula &&&&&&&&&&&& \\	
		&$X_1$ to $Y$&0.349&-0.054&0.295&0.388&-0.052&0.336&0.364&-0.072&0.292&0.368&-0.083&0.285  \\
		&$X_2$ to $Y$&0.363&-0.015&0.348&0.326&-0.023&0.303&0.284&-0.048&0.236&0.312&-0.074&0.238  \\
		&$X_3$ to $Y$&-0.335&0.115&-0.220&-0.292&0.080&-0.212&-0.283&0.070&-0.213&-0.281&0.091&-0.190 \\
		\hline
		Medium&\multicolumn{1}{c}{Classical}& \multicolumn{1}{c}{} & \multicolumn{1}{c}{} & \multicolumn{1}{c}{}&&&&&&&&&\\ 		
		&$X_1$ to $Y$&0.450&0.087&0.537&0.446&0.077&0.523&0.415&0.081&0.496&0.415&0.068&0.483  \\
		
		&$X_2$ to $Y$&0.301&0.135&0.436&0.285&0.131&0.416&0.292&0.125&0.417&0.286&0.118&0.404 \\
		\parbox[t]{5mm}{\multirow{1}{*}{\rotatebox[origin=c]{90}{Train-Set }}}
		&$X_3$ to $Y$&0.106&0.107&0.213&0.215&0.100&0.315&0.225&0.108&0.333&0.212&0.121&0.333 \\
		
		&Copula & & &&&&&&&&&& \\
		&$X_1$ to $Y$&0.449&0.087&0.536&0.447&0.077&0.524&0.415&0.080&0.495&0.415&0.069&0.484  \\
		&$X_2$ to $Y$&0.298&0.135&0.433&0.284&0.131&0.415&0.292&0.124&0.416&0.286&0.118&0.404 \\
		&$X_3$ to $Y$&0.106&0.107&0.213&0.216&0.100&0.316&0.225&0.107&0.332&0.211&0.122&0.333 \\
		
		%&\multicolumn{1}{c}{}& \multicolumn{1}{c}{} & \multicolumn{1}{c}{} & \multicolumn{1}{c}{}\\ 	
		Medium&Classical &&&&&&&&&&&& \\
		&$X_1$ to $Y$&0.450&0.082&0.532&0.446&0.070&0.516&0.415&0.081&0.496&0.415&0.067&0.482 \\
		
		&$X_2$ to $Y$&0.302&0.126&0.428&0.285&0.123&0.408&0.292&0.125&0.417&0.286&0.116&0.402 \\
		\parbox[t]{5mm}{\multirow{1}{*}{\rotatebox[origin=c]{90}{Test-Set }}}
		&$X_3$ to $Y$&0.106&0.116&0.222&0.215&0.104&0.319&0.225&0.111&0.336&0.212&0.121&0.333 \\
		&Copula &&&&&&&&&&&& \\	
		&$X_1$ to $Y$&0.443&0.084&0.527&0.425&0.068&0.493&0.419&0.078&0.497&0.419&0.067&0.486  \\
		&$X_2$ to $Y$&0.294&0.131&0.425&0.269&0.121&0.390&0.277&0.128&0.405&0.282&0.117&0.399  \\
		&$X_3$ to $Y$&0.134&0.115&0.249&0.224&0.099&0.323&0.232&0.107&0.339&0.214&0.121&0.335 \\
		\hline
	\end{tabular}\label{M4Table6}
\end{table}
\clearpage
\newpage
\begin{table}[h!]
	\caption{Summary of direct effect (D.E.), indirect effect (I.E.), and total effect (T.E.) obtained by high  correlation for $p=3$.} % title of Table
	\centering
	%	\footnotesize
	%		\scriptsize
	\tiny	
	\begin{tabular}{cccccccccccccc} % centered columns (6 columns)
		\toprule 
		Sample $\rightarrow$& \multicolumn{3}{c}{\hspace{2.5cm}$n = 100$}& &\multicolumn{2}{c}{\hspace{0.5cm}$n = 200$}&& \multicolumn{2}{c}{\hspace{0.7cm}$n = 300$} &&\multicolumn{2}{c}{\hspace{0.7cm}$n = 400$}&\\
		\cmidrule{3-5} \cmidrule{6-8} \cmidrule{9-11} \cmidrule{12-14}	
		%		\hline
		%		\hline
		%		Sample $\rightarrow$& &n=100&&&n=200&&&n=300&&&n=400&&\\
		$\downarrow$ Corr.&Approach&D.E. &I.E.&T.E.&D.E. &I.E.&T.E.&D.E. &I.E.&T.E.&D.E. &I.E.&T.E. \\ [1ex] % inserts table
		\hline
		High&\multicolumn{1}{c}{Classical}& \multicolumn{1}{c}{} & \multicolumn{1}{c}{} & \multicolumn{1}{c}{}&&&&&&&&&\\ 
		&$X_1$ to $Y$&0.605&0.189&0.794&0.436&0.260&0.696&0.390&0.279&0.669&0.403&0.273&0.676  \\
		&$X_2$ to $Y$&0.206&0.386&0.592&0.245&0.319&0.564&0.275&0.312&0.587&0.257&0.315&0.572  \\
		\parbox[t]{5mm}{\multirow{1}{*}{\rotatebox[origin=c]{90}{Train-Set}}}
		&$X_3$ to $Y$&0.140&0.412&0.552&0.271&0.294&0.565&0.291&0.303&0.594&0.303&0.309&0.612  \\	
		&Copula & & &&&&&&&&&& \\
		&$X_1$ to $Y$&0.605&0.189&0.794&0.436&0.259&0.695&0.389&0.279&0.668&0.402&0.274&0.676 \\
		&$X_2$ to $Y$&0.205&0.386&0.591&0.245&0.319&0.564&0.275&0.311&0.586&0.257&0.315&0.572  \\
		&$X_3$ to $Y$&0.140&0.411&0.551&0.272&0.293&0.565&0.291&0.302&0.593&0.303&0.309&0.612  \\
		\hline
		%&\multicolumn{1}{c}{}& \multicolumn{1}{c}{} & \multicolumn{1}{c}{} & \multicolumn{1}{c}{}\\ 	
		High&Classical &&&&&&&&&&&& \\
		&$X_1$ to $Y$&0.605&0.187&0.792&0.436&0.265&0.701&0.390&0.286&0.676&0.403&0.278&0.681\\
		&$X_2$ to $Y$&0.206&0.368&0.574&0.245&0.323&0.568&0.275& 0.321&0.596&0.257&0.325&0.582  \\
		\parbox[t]{5mm}{\multirow{1}{*}{\rotatebox[origin=c]{90}{Test-Set }}}
		&$X_3$ to $Y$&0.140&0.438&0.578&0.272&0.290&0.562&0.291&0.306&0.597&0.303&0.314&0.617  \\
		&Copula &&&&&&&&&&&& \\	
		&$X_1$ to $Y$&0.637&0.145&0.782&0.446&0.252&0.698&0.390&0.282&0.672&0.399&0.279&0.678 \\
		&$X_2$ to $Y$&0.156&0.373&0.529&0.223&0.328&0.551&0.271&0.321&0.592&0.255&0.324&0.579  \\
		&$X_3$ to $Y$&0.112&0.436&0.548&0.272&0.288&0.560&0.289&0.305&0.594&0.307&0.310&0.617  \\
		\hline
	\end{tabular}\label{M4Table7}
	
\end{table}

\section{Real Applications}
\begin{figure}[h]
	\centering
	\subfloat{\includegraphics[scale=0.56, angle=0]{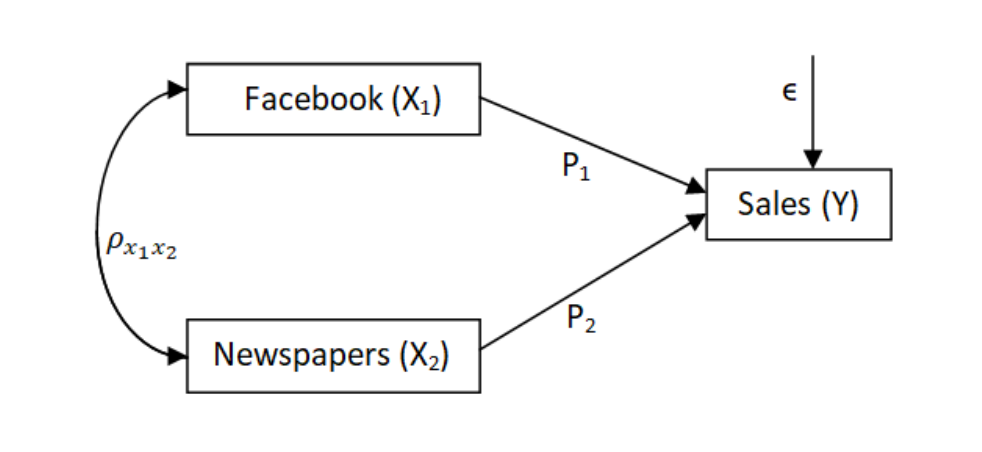}}
	\subfloat{\includegraphics[scale=0.65, angle=0]{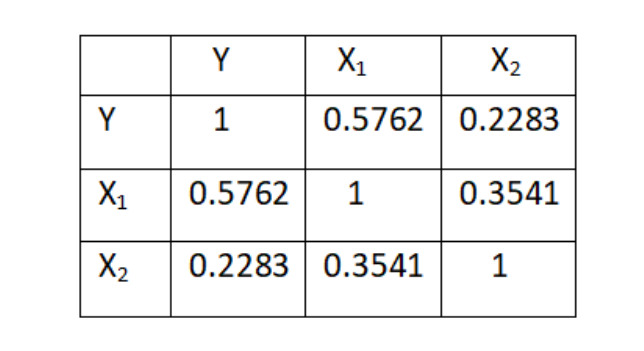}}
	\caption{ Path model (left) and correlation structure of variables (right).}
	\label{M4Fig.2}
\end{figure}

\begin{figure}[h]
	\centering
	\subfloat{\includegraphics[scale=0.56, angle=0]{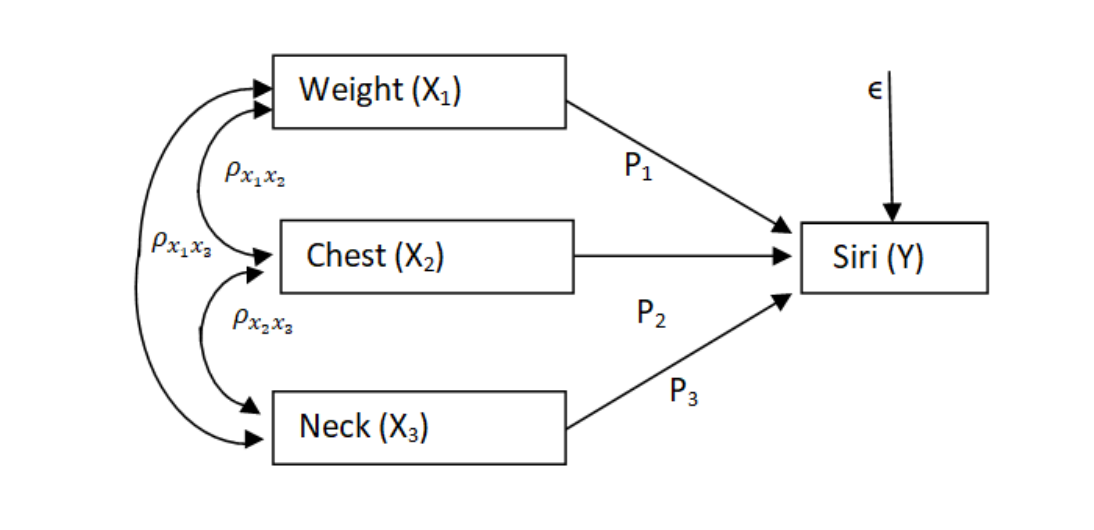}}
	\subfloat{\includegraphics[scale=0.63, angle=0]{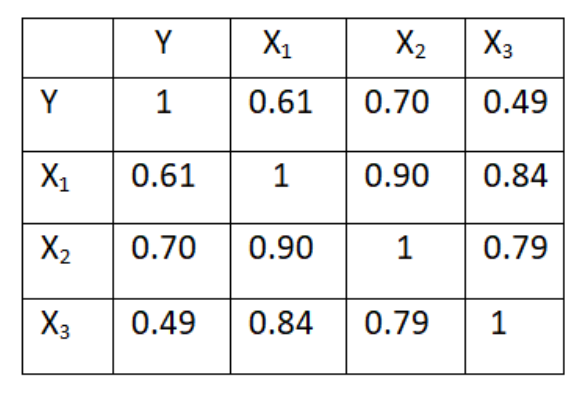}}
	\caption{ Path model (left) and correlation matrix (right).}
	\label{M4Fig.3}
\end{figure}

\noindent \textbf{Real Application 1:} Here, we consider the well known `marketing' data which is easily available in `datarium' package of R-software. It contains 200 observations with three advertising medias facebook, newspaper, and youtube on sales. In present study, we consider sales as a endogenous variable and facebook and newspaper as a exogenous variables, denoted it by $Y$, $X_1$, and $X_2$, respectively. In general, the expending money on advertising through facebook has better sales than newspapers but there are also some indirect effect on sales by advertising through newspapers. We use here classical and copula-based regression approaches to calculate the path coefficients and predict the significant direct and indirect effect of advertising medias facebook ($X_1$) and newspaper ($X_2$) on sales ($Y$).  First, we extracted each selected variable from the data frame and standardized it, then confirmed the normality using `fitur' package in R software. The p-values corresponding to Kolmogorov-Samirov test equal to 0.053, 0.119, and 0.051 suggest that standardized  $Y$, $X_1$, and $X_2$ support to the normal distribution. Path model and observed correlation structure of standardized variables are reflected in Fig \ref{M4Fig.2}. 
%Moreover, using `VineCopula' package, we found that the considered variable $Y$, $X_1$, and $X_2$ are connected via Gaussian copula with correlation coefficient reflected in fig.\ref{M4Fig.2}. 
We perform our analysis using 5-fold cross validation technique. The calculated model evaluation indices; mean of MSE, standard deviation (SD) of MSE, AIC, and BIC for train set and test set are reported in Table \ref{M4Table8}. Model evaluation indices in train set is almost same in both approaches but significant differences in test set devotes that the copula based regression approach works good over classical approach.
\begin{table}[h]
	\caption {Summary of model evaluation indices of marketing data.} % title of Table
	\centering
	%		\footnotesize
	\scriptsize
	%\tiny
	\begin{tabular}{cccc} % centered columns (6 columns)	
		\hline
		&\multicolumn{1}{c}{Indices}&\multicolumn{1}{c}{Classical Approach}& \multicolumn{1}{c}{Copula Approach} \\ 
		\hline  
		&Mean of MSE &0.661&0.661\\
		\parbox[t]{1cm}{\multirow{1}{*}{\rotatebox[origin=c]{90}{Train-Set}}}
		&SD of MSE &0.048&0.044\\
		&	AIC &-100&-100 \\
		&BIC &-90&-90\\
		\hline 
		&Mean of MSE &0.693&0.684\\
		\parbox[t]{1cm}{\multirow{1}{*}{\rotatebox[origin=c]{90}{Test-Set}}}
		&SD of MSE &0.177&0.175\\
		&AIC &-485.32&-486.95 \\
		&BIC &-475.42&-477.05\\	
		\hline
	\end{tabular}
	
	\vspace{0.3cm}
	\begin{minipage}{13cm}
		\small
		Note: Train set and test set contains 160 and 40 observations, respectively.\\
	\end{minipage} 	
	\label{M4Table8} % is used to refer this table in the text
\end{table}
As well as, the analyzed direct, indirect, and total effect of advertising medias; facebook ($X_1$) and newspaper ($X_2$) on sales ($Y$) for train and test set through both approaches are presented in Table \ref{M4Table9}. The main observation we found that the total effect of $X_1$ on $Y$, and $X_2$ on $Y$ are approximately equal to the correlation coefficient between $X_1$ on $Y$, and $X_2$ on $Y$, respectively, which verify our theoretical concepts.  \\ 
\begin{table}[h]
	\caption{Summary of effects of path model based on marketing data with path estimating structure regression equation  $Y$ = $P_1X_1$ + $P_2X_2$. } % title of Table
	\centering
	%		\footnotesize
	\scriptsize
	%		\tiny
	
	\begin{tabular}{ccccc} % centered columns (6 columns)
		
		\hline
		&Approach&Direct Effect &Indirect Effect&Total Effect \\ [1ex] % inserts table
		\hline
		&\multicolumn{1}{c}{Classical Approach}& \multicolumn{1}{c}{} & \multicolumn{1}{c}{} & \multicolumn{1}{c}{}\\ 
		
		&$X_1$ to $Y$ &0.5665 &0.0096 &0.5761 \\
		\parbox[t]{5mm}{\multirow{1}{*}{\rotatebox[origin=c]{90}{Train-Set }}}
		&$X_2$ to $Y$ & 0.0270 &0.2010&0.2280\\
		&Copula Approach & & & \\
		&$X_1$ to $Y$ &0.5665 & 0.0096  &0.5761 \\
		&$X_2$ to $Y$ & 0.0271 &0.2010&0.2281 \\
		\hline
		%&\multicolumn{1}{c}{}& \multicolumn{1}{c}{} & \multicolumn{1}{c}{} & \multicolumn{1}{c}{}\\ 	
		&Classical Approach &  && \\
		
		&$X_1$ to $Y$ &0.5665 &0.0098 &0.5763 \\
		\parbox[t]{5mm}{\multirow{1}{*}{\rotatebox[origin=c]{90}{Test-Set }}}
		&$X_2$ to $Y$ & 0.0270&0.2065&0.2335\\
		&Copula Approach & & & \\	
		&$X_1$ to $Y$ &0.5715 & 0.0044  &0.5759 \\
		&$X_2$ to $Y$ & 0.0122 &0.2084&0.2206 \\
		\hline
	\end{tabular}
	
	\vspace{0.3cm}
	\begin{minipage}{10cm}
		\small
		Note: (i) Total effect = Direct Effect + Indirect Effect.\\
		(ii) In train set  $\hat{\rho}_{x_1x_2}= 0.3548$ and in test set  $\hat{\rho}_{x_1x_2}= 0.3646$.
	\end{minipage} 	
	\label{M4Table9} % is used to refer this table in the text
\end{table}
%\clearpage
\newpage
\noindent \textbf{Real Application 2:}  Here, we consider the `bodyfat' data set reposted in Penrose et al. \cite{penrose1985generalized}. It contains the percentage of body fat as a dependent variable along with other covariates which represent several physiologic measurements related to 252 men. This data set is also easily available in R-software under package `mfp'.  In present study, we restrict ourself  on only four variables namely, siri (body fat in percent), weight (in lbs), chest circumference (in cm),  and neck circumference (in cm)  from the `bodyfat' data set. Here, we consider siri as a endogenous variable ($Y$) and   weight ($X_1$) , chest circumference ($X_2$), and neck circumference ($X_3$) as a exogenous variables. Path model and correlation structure of standardized variables are reflected in Fig \ref{M4Fig.3}.
\noindent We apply the same procedure to fit the model and obtained the model evaluation indices, direct, indirect, and total effect of variables for train and test set. All results of analyzed path model are reported in Table \ref{M4Table10} and Table \ref{M4Table11}, respectively. 
\begin{table}
	\caption{Summary of model evaluation Indices of bodyfat data.} % title of Table
	\centering
	%		\footnotesize
	\scriptsize
	%\tiny
	\begin{tabular}{cccc} % centered columns (6 columns)	
		\hline
		&\multicolumn{1}{c}{Indices}&\multicolumn{1}{c}{Classical Approach}& \multicolumn{1}{c}{Copula Approach} \\ 
		\hline  
		&Mean of MSE &0.4940&0.4942\\
		\parbox[t]{1cm}{\multirow{1}{*}{\rotatebox[origin=c]{90}{Train-Set}}}
		&SD of MSE &0.0332&0.0330\\
		&	AIC &-201&-201 \\
		&BIC &-180&-180\\
		\hline 
		&Mean of MSE &0.4990&0.4884\\
		\parbox[t]{1cm}{\multirow{1}{*}{\rotatebox[origin=c]{90}{Test-Set}}}
		&SD of MSE &0.1347&0.1305\\
		&AIC &-640&-645 \\
		&BIC &-619&-624\\	
		\hline
	\end{tabular}
	
	\vspace{0.3cm}
	\begin{minipage}{13cm}
		\small
		Note: Train set and test set contains 202 and 50 observations, respectively.\\
	\end{minipage} 	
	\label{M4Table10} % is used to refer this table in the text
\end{table}
%	\clearpage
%	\newpage
\begin{table}[h!]
	\caption{Summary of effects of path model based on bodyfat data with path estimating structure regression equation $Y$ = $P_1X_1$ + $P_2X_2$ + $P_3X_3$ . } % title of Table
	\centering
	%		\footnotesize
	\scriptsize
	%\tiny
	
	\begin{tabular}{ccccc} % centered columns (6 columns)
		
		\hline
		&Approach&Direct Effect &Indirect Effect&Total Effect \\ [1ex] % inserts table
		\hline
		&\multicolumn{1}{c}{Classical Approach}& \multicolumn{1}{c}{} & \multicolumn{1}{c}{} & \multicolumn{1}{c}{}\\ 
		
		&$X_1$ to $Y$ &0.0294&0.5831&0.6125 \\
		
		&$X_2$ to $Y$ &0.8078 &-0.1055&0.7023\\
		\parbox[t]{5mm}{\multirow{1}{*}{\rotatebox[origin=c]{90}{Train-Set }}}
		&$X_3$ to $Y$ &-0.168&0.6582&0.4902 \\
		&Copula Approach & & & \\
		&$X_1$ to $Y$ &0.0296&0.5833&0.6129 \\
		&$X_2$ to $Y$ &0.8079&-0.1053&0.7026 \\
		&$X_3$ to $Y$ &-0.1679&0.6584&0.4905 \\
		\hline
		%&\multicolumn{1}{c}{}& \multicolumn{1}{c}{} & \multicolumn{1}{c}{} & \multicolumn{1}{c}{}\\ 	
		&Classical Approach &  && \\
		
		&$X_1$ to $Y$ &0.0294&0.5860&0.6154 \\
		
		&$X_2$ to $Y$ &0.8078 &-0.1050&0.7028\\
		\parbox[t]{5mm}{\multirow{1}{*}{\rotatebox[origin=c]{90}{Test-Set}}}
		&$X_3$ to $Y$ &-0.168&0.6557&0.4877 \\
		&Copula Approach & & & \\	
		&$X_1$ to $Y$ &0.041&0.5758&0.6168 \\
		&$X_2$ to $Y$ &0.7894&-0.0886&0.7008 \\
		&$X_3$ to $Y$ &-0.1603&0.6509&0.4906 \\
		\hline
	\end{tabular}
	
	\vspace{0.3cm}
	\begin{minipage}{14cm}
		\small
		Note: (i) Total effect = Direct Effect + Indirect Effect.\\
		(ii) In train set $\hat{\rho}_{x_1 x_2}=0.8945 $, $\hat{\rho}_{x_1 x_3}=0.8301 $, $\hat{\rho}_{x_2x_3}=0.7846 $, and in test set  $\hat{\rho}_{x_1 x_2}=0.8967 $, $\hat{\rho}_{x_1x_3}=0.8234 $, $\hat{\rho}_{x_2x_3}=0.7818 $.
	\end{minipage} 	
	\label{M4Table11} % is used to refer this table in the text
\end{table}
From Table \ref{M4Table10}, it can be observe that the model evaluation indices; mean of MSE, SD of MSE, AIC, and BIC for train  set are almost equal but significance difference in test set supports the superiority of copula approach.  We found the same interpretation of other results for this data also.	

%\clearpage
%\newpage
\section{Conclusion}
\noindent The present study explored the path analysis models when the endogenous (dependent) variable and exogenous (independent) variables are jointly associated with the elliptical copulas and  investigated the efficacy of path models when direct and indirect effects are estimated using classical ordinary least squares  and copula-based regression approaches in different scenarios. First, we developed the basic structure of the regression models in terms of copulas and explored some typical examples for the Gaussian and $t$-copulas. Based on the well-organized numerical scheme, we estimate the path coefficients using classical and copula approaches and also study the direct and indirect effects of exogenous variables in path models. This study suggest that the copula-based path analysis performs better over the classical approach in terms of mean of MSE, standard deviation of MSE, AIC, and BIC. In the copula approach, we observed that the total effect of each exogenous variable on the endogenous variable is approximately equal to their correlation coefficient. Apart from the present study, we intend to explore the distribution of the indirect effects using copula approach in near future.\\
\noindent \textbf{Disclosure statement:}\\
No potential conflict of interest was reported by the author(s).

\bibliographystyle{abbrv}
\bibliography{Path_bib}

\end{document}